\documentclass[]{article}
\usepackage{arxiv}
\usepackage{amsmath,amssymb,amsfonts} 
\usepackage{epsfig} \usepackage{latexsym,nicefrac,bbm}
\usepackage{xspace}
\usepackage{color,fancybox,graphicx,fullpage}
\usepackage{tabularx}
\usepackage{hyperref} 
\usepackage{pdfsync}
\usepackage[boxruled]{algorithm2e}
\usepackage{multicol}
\usepackage[inline]{enumitem}
\usepackage{multirow}
\usepackage{url}
\usepackage{bm}
\usepackage{bbm}
\usepackage{caption}
\usepackage{subcaption}

\usepackage{listings}
\usepackage{makecell}
\usepackage{booktabs}
\usepackage{amsmath,bm}
\usepackage{multirow}

\usepackage{multibib}

\begin{document}
\title{On hate scaling laws for data-swamps}
\author {
    Abeba Birhane\thanks{Equal contribution} \\
    Mozilla Foundation, San Francisco, USA \& \\
 School of Computer Science and Statistics\\
    Trinity College Dublin, Ireland \\
    \texttt{birhanea@tcd.ie} \\ \And 
    Vinay Prabhu\textsuperscript{*}, Sang Han\\
    Independent researchers  \\
  San Francisco\\
    USA\\
    \texttt{vinaypra@alumni.cmu.edu,sanghan@protonmail.com}\\ \And
    Vishnu Naresh Boddeti \\
    Computer Science and Engineering \\
    Michigan State University \\
    \texttt{vishnu@msu.edu}
    }
\date{}

\maketitle

\begin{abstract}
`Scale the model, scale the data, scale the GPU-farms' is the reigning sentiment in the world of generative AI today. While model scaling has been extensively studied, data scaling and its downstream impacts remain under explored. 
This is especially of critical importance 
in the context of visio-linguistic datasets whose main 
source 
is the World Wide Web, condensed and packaged as the \textit{CommonCrawl} dump. 
This large scale data-dump, which is known to have numerous drawbacks, 
is repeatedly 
mined and serves as the data-motherlode for large generative models. In this paper, we: 1) investigate the effect of scaling datasets on hateful content through a comparative audit of 
the LAION-400M and LAION-2B-en, containing 400 million and 2 billion samples respectively, and 2) evaluate the downstream impact of scale on visio-linguistic models trained on these dataset variants by  measuring racial bias of the models trained on them using the Chicago Face Dataset (CFD) as a probe. 
Our results show that 1) the presence of hateful content in datasets, when measured with a Hate Content Rate (HCR) metric on the inferences of the Pysentimiento hate-detection Natural Language Processing (NLP) model, \textit{increased by nearly $12\%$} and 2) societal biases and negative stereotypes were also exacerbated with scale on the models we evaluated. 
As scale increased, the tendency of the model to associate images of human faces with the `human being' class over 7 other offensive classes \textit{reduced by half}. 
Furthermore, for the Black female category, the tendency of the model to associate their faces with the \textit{`criminal'} class \textit{doubled}, while \textit{quintupling} 
for Black male faces. 
We present a qualitative and historical analysis of the model audit results, reflect on our findings and its implications for dataset curation practice, and close with a summary of our findings and potential future work to be done in this area.    
All the meta-datasets curated in this endeavor and the code used are shared at: \url{https://github.com/vinayprabhu/hate_scaling}.

{\color{red}\textit{Content warning: This article contains examples of hateful text and NSFW images that might be disturbing, distressing, and/or offensive.}}
\end{abstract}

\maketitle

\section{Introduction}%

\label{sec:intro}
Generative AI models have come to captivate diverse stakeholders, spanning from researchers~\cite{birhane2021multimodal, luccioni2021s, koch2021reduced}, to media institutions~\cite{GenAI_2_MIT,GenAI_3_ars}, and even large-scale investment firms~\cite{GenAI_4_GS,GenAI_1_a16z}. %
This trend can be traced back to the emergence of Dall.E~\cite{dallE_ramesh2021zero}, a text-to-image visio-linguistic model released in April 2022, which purportedly attracted over a million users within the first three months of its launch, and was celebrated with claims like: “[t]he first AI technology that has caught fire with regular people”~\cite{GenAI_2_MIT}. Subsequently, models such as StableDiffusion~\cite{rombach2022high} and \href{https://www.midjourney.com/}{Midjourney} emerged, followed by black box projects from Big Tech such as Imagen~\cite{saharia2022photorealistic_imagen}, Parti~\cite{yu2022scaling_parti}, and BASIC~\cite{pham2021combined_basic}, access to which was never given to the general public. While Stable Diffusion and its variants have been trained on the open-sourced datasets from the LAION family%
 , little is known about the datasets that were used to train models such as Dall-E, Parti~\cite{yu2022scaling_parti}, and Imagen~\cite{saharia2022photorealistic_imagen}.

Fundamental to this %
multimodal model boom is large-scale visio-linguistic datasets containing image-text pairs, which form the main focal point of this paper. Broadly speaking, these datasets are of two types: those that are open-source, ``freely available'' and mainly scraped from the \href{https://commoncrawl.org}{Common Crawl} (such as LAION-400M~\cite{schuhmann2021laion400m} and LAION-5B~\cite{schuhmann2022laion5b}), and those that are closed %
datasets curated internally by Big Tech corporate labs (such as Google's ALIGN 1.7B/ALIGN 6.6B~\cite{jia2021scaling_align}, JFT-5B~\cite{pham2021combined_basic}, and OpenAI's WebImageText-WIT~\cite{clip_radford2021learning}). The latter remain %
outside the %
reach of independent audits and evaluations, while the models trained on such datasets are public-facing and commercialized via %
APIs (such as Microsoft Bing image-creator powered by Dall.E, or the \href{https://platform.openai.com/docs/guides/images}{Dall.E API}). %
These models are also being adopted in various commercial tools and applications such as stock photo generation~\cite{shutterstockdalle2022}, which contribute to accelerating their adoption and usage.

The open-source variants of these datasets are getting bigger and now breaching the billion-samples mark for two reasons: firstly, there is the unquestioned subservience to the \emph{scale is all you need} mandate handed down from Big Tech disseminations~\cite{jia2021scaling_align,birhane2021multimodal} that forms the motivational drive. Secondly, there is the emergent nexus between dataset curation and venture capital resulting in capital infusion into these dataset curation efforts which was hitherto missing: the LAION-5B~\cite{schuhmann2022laion5b} dataset, for example, was sponsored by Hugging Face, Doodlebot and Stability.ai, as per their \href{https://laion.ai/blog/laion-5b/}{blog-post announcement}.

In turn, this \emph{scale is all you need} mandate emerges from two schools of reasoning in published literature venerating scale.
The first pertains to vague ``dataset scaling laws" that we cover in detail in Appendix~\ref{app:dataset_scaling_laws} and the second pertains to the non-reproducible empirical results buried in subsections of some of the canonical papers which describe the (closed) datasets used for training models such as ALIGN~\cite{jia2021scaling_align}, Imagen~\cite{saharia2022photorealistic_imagen} and Parti~\cite{yu2022scaling_parti} (See Appendix~\ref{app:dataset_scaling_results}). We also note that these high-profile disseminations are increasingly turning into a flag-posting exercise that involves tactfully concealing critical details on the manner in which the dataset was curated and where the data came from (See Appendix~\ref{app:tactical_template} for a deeper exploration).

All of this has resulted in a \textit{``scrape-first-ask-questions-later"} data creation and curation culture, generating plausibly illegal gargantuan datasets and models, that has in turn elicited a slew of 
copyright lawsuits~\cite{StableDi11:online_lawsuit}, en masse fetishization of women's bodies in an emergent synthetic digital culture~\cite{StableDi72:online_fetishization}, outright bans of model outputs from art-forums~\cite{Floodedw12:online_ban}, and marquee 
datasets filled with %
duplicates~\cite{webster2023duplication}. These dataset and model concerns and drawbacks, in turn, result in downstream negative impacts, often against marginalized groups, for example: exacerbation of negative stereotypes and biases~\cite{luccioni2023stable, garcia2023uncurated,birhane2021multimodal}, discriminatory and harmful representation~\cite{tomasev2022manifestations,bender2021dangers,weidinger2021ethical} and cultural and linguistic homogeneity~\cite{basu2023inspecting,bender2020climbing,blodgett2020language}. 

Hateful, abusive, racist, aggressive and targeted speech 
are overlapping phenomena %
yet each can be characterized along dimensions such as directed, generalized, explicit and implicit abuse~\cite{waseem2017understanding}. Oftentimes, the common targets of hateful speech are minoritized groups. Based on analysis of generated data to improve hate detection,~\cite{vidgen2020learning} highlight that most of the common targets of hate include Black people, women, Muslims and trans people. Furthermore, hateful, abusive, and aggressive speech is a systemic problem. %
~\cite{davidson2019racial}, for example, examined hate speech and abusive language detection datasets and found systematic racial bias in all datasets. Subsequently, they %
found that classifiers trained on them predict tweets written in African-American English as abusive at a substantially higher rate. Similarly, ~\cite{abid2021persistent} studied the outputs generated from GPT-3  when the word ``Muslim'' is included in the prompt. They found that 66 out of the 100 completions are violent, where these %
violent completions are less likely for other religions. While issues concerning hate speech are rooted in systemic structures, current attention has focused too much on %
finding toxic language, performance maximization and engineering solutions. ~\cite{prabhakaran2020online}, thus argue more attention ought to be paid to %
recognizing the root causes and focus on the social and ethical initiatives along with the real-world impacts of hate-speech.

In this paper, we examine:
\begin{enumerate*}[label=\arabic*)]
    \item %
the impact of scale on hate-speech through audits of %
textual descriptions in two datasets: LAION-400M and LAION-2B-en, and 
    \item the downstream negative impact on models trained on these two datasets through audits of such models trained on their variants. %
\end{enumerate*}

The rest of the paper is organized as follows. 
In Section~\ref{sec:survey}, we survey %
previous work on scale within the broader technical landscape as well as within the %
Machine Learning (ML) community. 
In Section~\ref{sec:alt_text}, we present
our dataset audit %
methodology followed by our findings in Section~\ref{sec:nlp_results} which reveal that hateful content increased when the dataset size was scaled from 400 million to 2 billion. %
To establish this, we used an NLP-aided quality audit of the datasets by measuring the hate content in the alt-text image descriptions using a state-of-the-art (SoTA) pre-trained open-source model named \texttt{\textit{Pysentimiento}}~\cite{perez2021pysentimiento}. We then focus on %
the downstream consequences of the hate-scaling phenomenon by measuring the racial biases 
exhibited by visio-linguistic models that have been trained on these two datasets, where we detail our %
 audit methodology and experimental design in Section~\ref{sec:cfd_experiments}. 
Our findings, summarized in Section~\ref{sec:cfd_results}, %
demonstrate that %
associations of Black people's faces with %
dehumanizing classes such as \texttt{criminals} markedly increased when the model size and architecture are held constant  and the dataset is scaled from 400 million to 2 billion. We delve deeper and position our experimental findings in historical context by qualitatively examining historical patterns of dehumanization and criminalization of Black bodies in Section~\ref{sec:dehuman}. Section~\ref{sec:recommendations} provides some caveats, discussions, and recommendations for equitable, responsible and accountable dataset creation, curation, and management practice. We then highlight a number of extensions for future work and conclude in %
Section~\ref{sec:conclusion}.
\section{Scale: An Overview \label{sec:survey}}

Current thinking around scale can be broadly categorized under two differing approaches:  
that which sees scale as a solution to problems such as model performance and generalization, and that which emphasizes numerous concerns that arise with an unwavering commitment to scale. We present both below. 

\subsection{Scale as a Solution}
The race to scale is a fixation driving not only research in ML but also the larger tech ``innovation'' discourse. Entrepreneurs are warned that \textit{`‘if you don’t know how to scale, don’t innovate’'}~\cite{pfotenhauer2022politics,seelos2017innovation}. %
Marked by the taken for granted, field-wide concept of \textit{scaling laws}, large scale is thought to correlate with better model performance in ML~\cite{kaplan2020scaling}. In fact, model performance, according to Schumann et al.~\cite{schuhmann2021laion}, can be improved by scaling up datasets, while Birhane et al. found that ``scaling up'' is one of the top desired values in ML research amongst the top 100 most influential ML papers of the past decade published in two of the most prestigious AI conferences (NeurIPS and ICML)\cite{birhane2021values}.

Scale is furthermore presented as a %
shortcut that can circumvent 
various dataset  curation related problems such as problematic content, resource-intensive dataset curation, and costly annotation processes, where larger scale is seen as a substitute for quality data and to ensure coverage of long tail of ``uncommon'' samples. Jia et al., for example, claim that: \textit{``heavy work on data curation and annotation''} can be avoided by scaling up image-text datasets~\cite{jia2021scaling_align}. The ``scale beats noise'' narrative has tactfully re-framed %
thoughtful 
handheld dataset curation %
as a costly %
problem that can be ``solved'' %
by larger scale. Scale, according to such narrative, is %
a liberating panacea that not only frees the downstream ML pipeline from the burdens of expensive filtering or post-processing steps but also makes up for ``noisy'' data %
as if  captioning errors in multimodal datasets of image and alt-text pairs can somehow be averaged out through the correct captioning elsewhere in the dataset. Such lines of thinking are not unique to this specific context, but form a widespread belief that drives initiatives such as the LAION datasets and permeate the entire field of the multi-modal models. %

\subsection{The Cost of Scale Thinking}

The primary motivation behind the LAION-400M undertaking was to produce open-source variants of the opaque Web-Image-Text (WIT) dataset, and the CLIP~\cite{clip_radford2021learning} and DALL.E~\cite{dallE_ramesh2021zero} models.
Such open-sourcing initiatives are important first steps towards accountability and building trustworthy AI given that for any auditing and evaluating to take place, open access is a crucial prerequisite. Nonetheless, given the numerous concerns %
that arise with web-sourced data, continual evaluation and audit of large-scale datasets and models is imperative for well-functioning, just, and healthy open-sourcing practices. 

For instance, Science and Technology Studies (STS) scholars and critical data and AI studies have repeatedly emphasized that ``scale thinking'' stands in stark opposition to values such as  societal equity or effective systemic change~\cite{hanna2020against, koch2021reduced}. In fact, unwavering commitment to scalability is instrumental to the realization of Big Tech's objectives, such as profit maximization and market monopoly, enabling the centralization of power in the handful few. This often comes at the expense of advancing and cultivating values such as individuals' rights, informed consent, justice, and consideration for societal impacts of models~\cite{birhane2021values}. 

With the awareness of the amplified negative downstream impacts of scale, there has also been increased attention towards the need to evaluate and audit models and large-scale datasets as an important intervention and accountability mechanism~\cite{raji2019actionable,metaxa2021auditing,vecchione2021algorithmic,luccioni2021s}. The recent emergence of grassroots-based open-sourcing initiatives come as a response to the increasing adoption of the closed-source commercial API access mode of dissemination being used for projects such as GPT-3~\cite{brown2020language}, CLIP, and DALL.E.
For instance, EleutherAI %
has
achieved success by replicating both the \textit{WebText} dataset (on which GPT-3 was trained) and the GPT-3 model itself by carefully curating and disseminating the Pile dataset~\cite{gao2020pile} and training and sharing the GPT-Neo~\cite{gpt-neo_1}/GPT-NeoX~\cite{gpt-neo_2} models. In this regard, the open-sourcing movement has been critical, enabling open access to datasets and models, which is key for independent auditing and evaluation.

\section{Dataset Audit: LAION-400M and LAION-2B-en}
\label{sec:alt_text}

One of the challenges of auditing multimodal datasets is that hateful content, negative stereotypes, and otherwise harmful and marginalizing representations can be present in either modality: text or image. This means that audits %
can span techniques ranging from image content analysis~\cite{somepalli2022diffusion_goldstein,birhane2021multimodal}, image source analysis (by analyzing the URL field), image-text cross-modal analysis (looking for discordance between an image and its alt-text description) and alt-text content %
analysis. Poor data quality, for example, is a common issue that arises with scale. Audits on image content analysis, for example, have revealed that  %
nearly $30\%$ (approximately 700 million image-text pairs) in the LAION-2B-en dataset are duplicates~\cite{webster2023duplication}. %
This, as addressed in \cite{somepalli2022diffusion_goldstein} and \cite{webster2023duplication}, can manifest as \textit{Digital Forgery}, or exact memorization of training examples present multiple times in training data, which was shown to be possible in recent work by Carlini et al~\cite{carlini2023extracting} -- a phenomenon that has stark ramifications for the field of image generation at large.

\subsection{Audit Methodology}
Our audits are focused on two versions of the LAION visio-linguistic datasets: LAION-400M~\cite{schuhmann2021laion400m} and LAION-2B-en, the English-language subset of the larger LAION-5B dataset~\cite{schuhmann2022laion5b} that consists of 2.32 billion image-text pairs.
The LAION-400M dataset is currently available as a collection of 32 randomly sampled subsets and are obtainable as individual parquet~\footnote{``Apache Parquet is an open source, column-oriented data file format designed for efficient data storage and retrieval. It provides efficient data compression and encoding schemes with enhanced performance to handle complex data in bulk''} %
files with a  mean size of 1.68 GB, for a total of 54 GB.  The LAION-2B-en, on the other hand, consists of 128 parquet files, each with a mean size of 2.52 GB (and a total size of 321 GB). Each of these parquet files contains image-text data pertaining to the following data fields: \emph{[`SAMPLE\_ID',  `URL', `TEXT', `HEIGHT', `WIDTH', `LICENSE', `NSFW', `similarity']}. 

\begin{figure*}
    \centering
    \includegraphics[width=0.6\columnwidth]{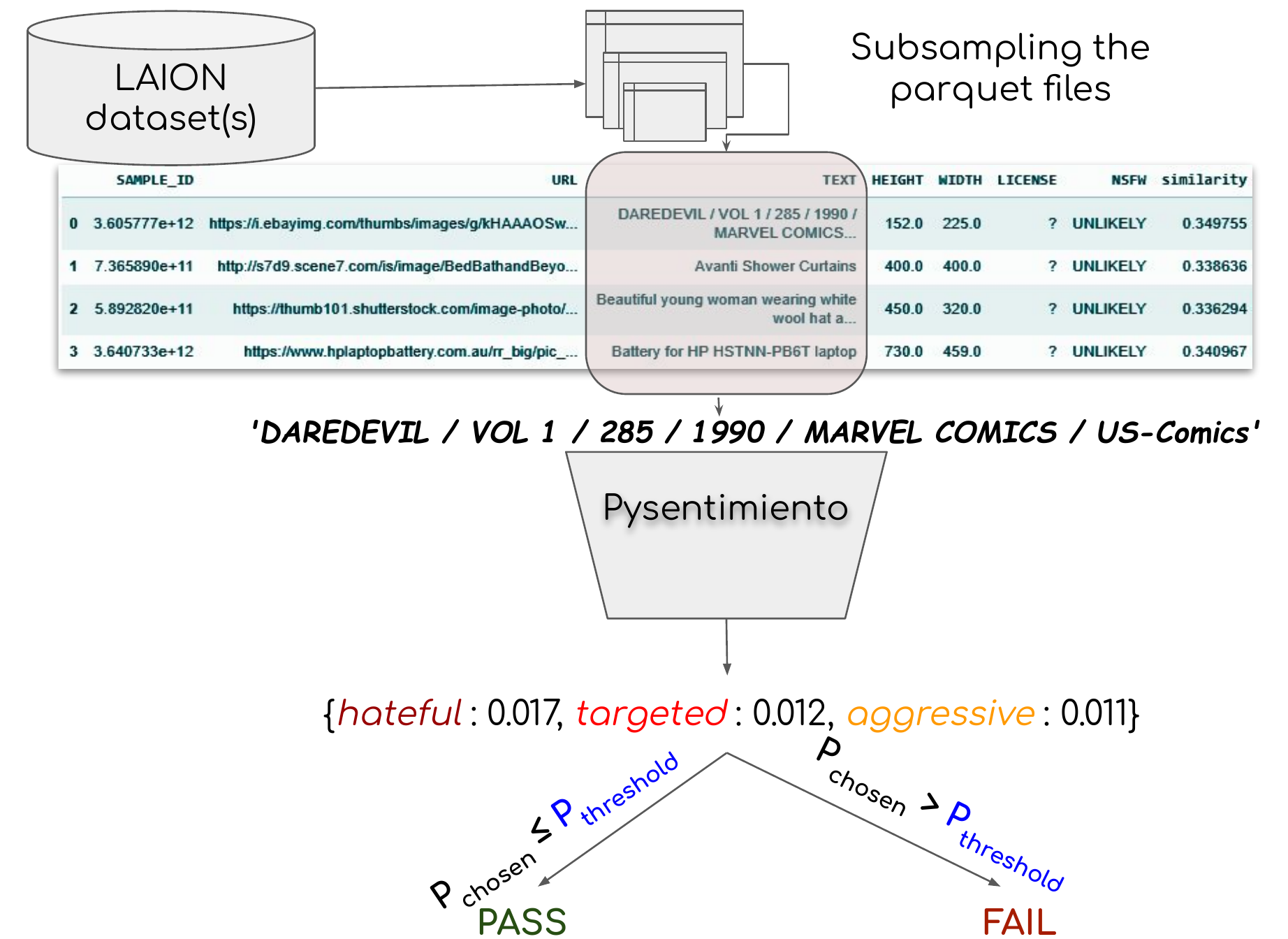}
    \caption{Experimentation details: Dataset sub-sampling, inference using Pysentimiento and thresholding for estimating Hate Content Rate (HCR).}
    \label{fig:exp_flow}
\end{figure*}

In order to evaluate the impact of scaling a dataset from 400 million to 2 billion samples on hateful content, %
we perform the following audits. %
We first sub-sample the dataset(s) and then extract the alt-text descriptions associated with the sampled image-rows in the \texttt{`TEXT'} field (See Figure~\ref{fig:exp_flow}). In all our experiments, we randomly sample 0.1 million rows from each of the $160\ (=32+128)$ constituent parquet files spanning the two datasets. This yields $N_{samples,400M}=3.2$ million samples for the LAION-400M dataset and $N_{samples,2B-en}=12.8$ million samples for the LAION-2B-en datasets respectively. To this end, we use \textit{Pysentimiento}~\cite{perez2021pysentimiento}, a SoTA open-source NLP framework.

To begin with, we define the metric, Hate Content Rate (HCR)~\footnote{We use the metric Hate Content Rate (HCR) as a shorthand for not just hateful content but all the three categories: hateful, targeted, and aggressive.}: $\psi_{type} \left( {{P_{threshold}}} \right)$ to be,
\begin{equation}
\psi_{type} \left( {{P_{threshold}}} \right) = 100 \times \frac{{\sum\limits_{i = 1}^{{N_{samples}}} {\mathbbm{1}\left(\tilde{p}_{type,i} > P_{threshold} \right)} }}{{{N_{samples}}}}
\end{equation}
where $\mathbbm{1}(\cdot)$ is the indicator function, ${\tilde p}_{type,i}$ is the probability score assigned by the Pysentimiento model for the text associated with the $i^{th}$  sample and $type \in \left\{ {'{\rm{hateful}}','{\rm{targeted}}','{\rm{aggressive}}'} \right\}$.

This captures the ratio of samples (as a percentage) that resulted in the Pysentimiento model assigning the associated hate/targeted/aggressive speech probability score to be greater than $P_{threshold}$.
We perform a comparative analysis of the extent of \textit{hate speech}, \textit{targeted speech} and \textit{aggressive speech} contained in them using Pysentimiento. %
In response to an input sentence, \textit{Pysentimiento} outputs a $3 \times 1$ vector containing probability scores across the three categories of \textit{hateful}, \textit{targeted} and \textit{aggressive} speech (see Table~\ref{tab:sample_alt_text} for randomly selected samples). The extracted alt-text descriptions are then passed through Pysentimiento to extract the $N_{samples} \times 3 $ text-quality score matrices for each of the two datasets.

\begin{table}
\caption{{Samples of alt text descriptions found in the dataset and the probability scores across the three categories of \textit{hateful}, \textit{targeted} and \textit{aggressive} speech.}\label{tab:sample_alt_text}}
\centering
\scalebox{0.9}{
\begin{tabular}{@{}p{0.78\textwidth}rrl@{}}
\toprule
{Alt text} &      {$P_{hateful}$} &    {$P_{targeted}$} &        {$P_{aggressive}$} \\
\midrule
`Biden's Spending Will Go To Illegal Immigrants While Tax Hikes Will Destroy American Jobs'   &  0.902 &  0.024 &  0.449 \\
`If you know this man, please, for the love of God tell him to BURN these pants!!'   &  0.401 &  0.262 &  0.517 \\
`shut up and be a don like nancy - Personalised Men's Long Sleeve T-Shirt' &   0.395 &  0.559 &  0.128\\
`This bored rich blonde shoplifter gets rough f**keds'  & 0.934 & 0.895 & 0.128\\
`Horny slave tied to tree gets pulled on her beautiful tits and gets hit on her c*nt with a stick and hands' & 0.983 & 0.911  & 0.909\\
\bottomrule
\end{tabular}}
\end{table}

We also introduce the \textit{‘Any-of-the-three’} detector that maps to the case where the input text fails the quality test if \textit{\textbf{any}} of $\tilde p_{hateful}$, $\tilde p_{aggressive}$ or $\tilde p_{targeted}$ happens to be greater than $P_{threshold}$. The associated \texttt{‘Any-of-the-three’}-HCR, $\bar \psi ({{P_{threshold}}})$ would be:
\begin{equation}
\begin{aligned}
\bar{\psi}\left(P_{threshold}\right) &= 100\times\frac{\sum\limits_{i = 1}^{N_{samples}} \mathbbm{1}\left(
(\tilde{p}_{hateful,i} > P_{threshold}) \mbox{ } \| \mbox{ } (\tilde{p}_{targeted,i} > P_{threshold}) \mbox{ } \| \mbox{ } (\tilde{p}_{aggressive,i} > P_{threshold})\right)}{N_{samples}}\\
&= 100 \times \frac{\sum\limits_{i=1}^{N_{samples}}\mathbbm{1}\left(\max \left\{\tilde{p}_{hateful,i},\tilde{p}_{targ eted,i},\tilde{p}_{aggressive,i} \right\}>P_{threshold}\right)}{N_{samples}}
\end{aligned}
\end{equation}

We then %
perform a quality check where we evaluate if one (or any) of the 3 probability score values associated with an input alt-text description exceeds a certain pre-set threshold score $P_{threshold}$, in which case the input text is deemed to have failed the quality check at that threshold (Figure~\ref{fig:exp_flow} illustrates this process).
We lastly compare the statistics associated with the text-quality score matrices to understand the nature of the text that was scooped in when the dataset expanded from 400 million samples to 2 billion samples.

\section{Dataset Audit Results}
\label{sec:nlp_results}
In this section, we use both HCR and, more specifically, `Any-of-the-three`-HCR, $\bar \psi ({{P_{threshold}=0.5}})$  as the default metric of comparison to characterize the amount of problematic content in both LAION-400M and LAION-2B-en datasets. We also and carry out a file-wise comparison of specific shards of both datasets. 

\subsection{Scaling is not Benign: Comparing LAION 400M and LAION 2B-en\label{subsec:sub-sampling_bad}}
\begin{figure}
    \centering
    \includegraphics[width=0.8\textwidth]{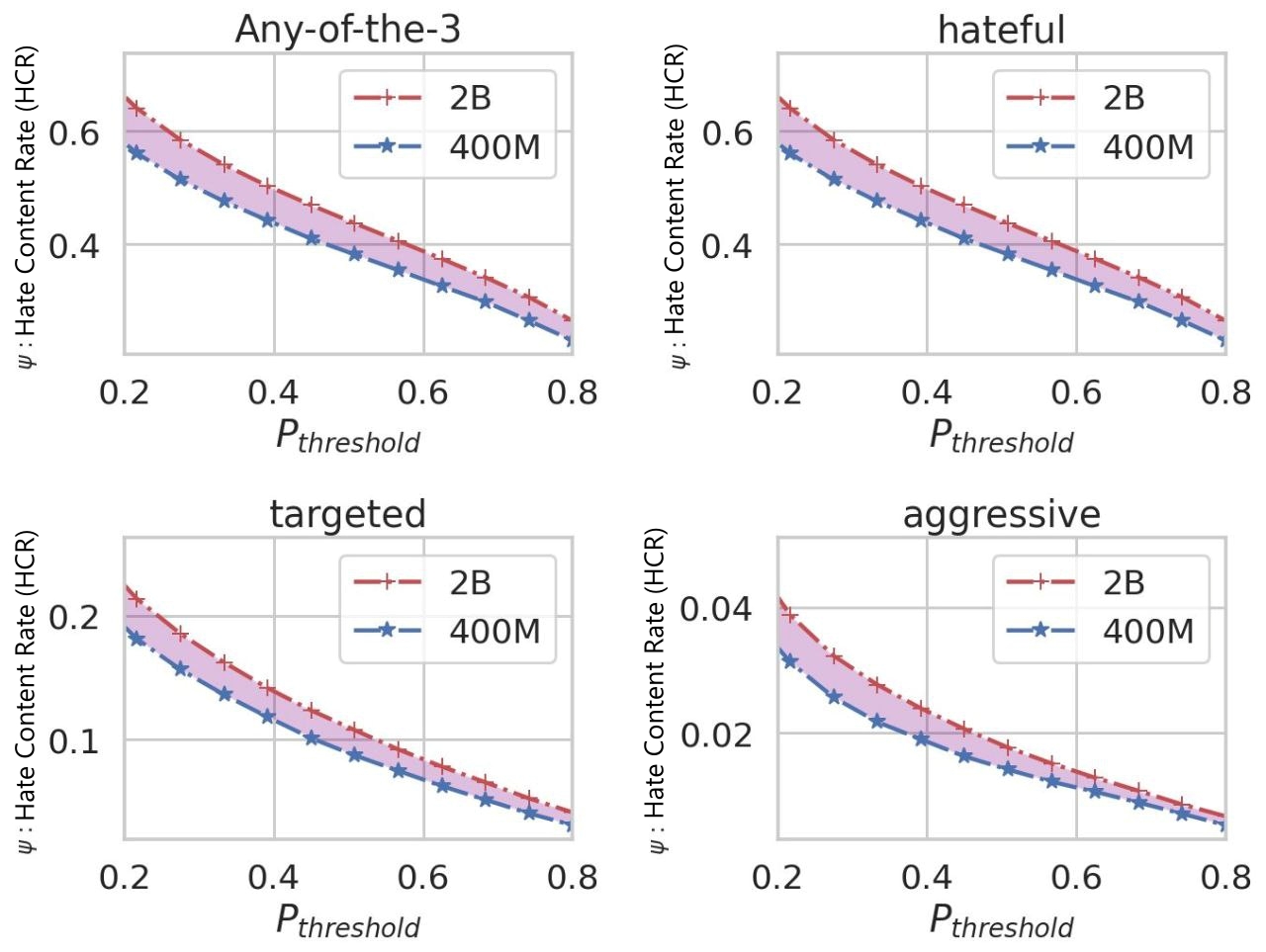}
    \caption{HCR curves for the LAION400M and LAION-2B-en datasets using \textit{Pysentimiento} outputs. %
    As the dataset is scaled, there is a statistically significant increase in hateful content.} %
    \label{fig:qfr_types}
\end{figure}

We begin by focusing on Figure~\ref{fig:qfr_types} that presents a plot of the HCR curves as a function of $P_{threshold}$. We observe that, as $P_{threshold}$ increases, the HCR curves monotonically decrease, indicating that fewer textual samples meet the more stringent constraint placed by a higher $P_{threshold}$ value. Worryingly however, we found that for all the sentiment types -- \textit{hate, targeted}, and \textit{aggressive }speech -- the HCR-curve(s) pertaining to the 2B-en dataset lies \textit{\textit{strictly above}} the 400M dataset’s curve(s). This signifies that irrespective of what $P_{threshold}$ is being chosen, the quality failure rate signifying the prevalence of hateful content %
is higher with the 2B-en dataset in comparison to its 400M counterpart.  
We found that amongst the three sentiment types, the 'hateful' type emerged as the most prevalent for both datasets, with the 2B dataset having a HCR of up to \textbf{0.7\%} and 400M one of \textbf{0.6\%}, followed by the 'targeted' type, with an HCR up to \textbf{0.25\% v/s 0.2\%}, and finally the 'aggressive' type, with an HCR of \textbf{0.04\% v/s 0.03\%}. 

For the \texttt{‘Any-of-the-three’} curve (leftmost in Figure~\ref{fig:qfr_types}), we observe that 
the HCR curves pertaining to the 2B-en dataset is above the the curves of the 400M dataset. 
Given that both LAION-400M and LAION-2B-en are extracted from the \textit{CommonCrawl} dataset, we hypothesize that during the race to expand the dataset to 2 billion samples, the dataset scraping module might have sampled from the \textit{low-quality sub-graphs} of the \textit{CommonCrawl} graph at a rate worse than that during the LAION-400M %
creation process. We also note that the \textit{CLIP-filtering} \href{https://laion.ai/blog/laion-5b/}{threshold} to have been relaxed from \textbf{0.3 }(for LAION-400M) to \textbf{0.28} (for LAION-2B-en), which could be another explanatory factor.

\begin{figure*} 
    \centering
    \includegraphics[width=\columnwidth]{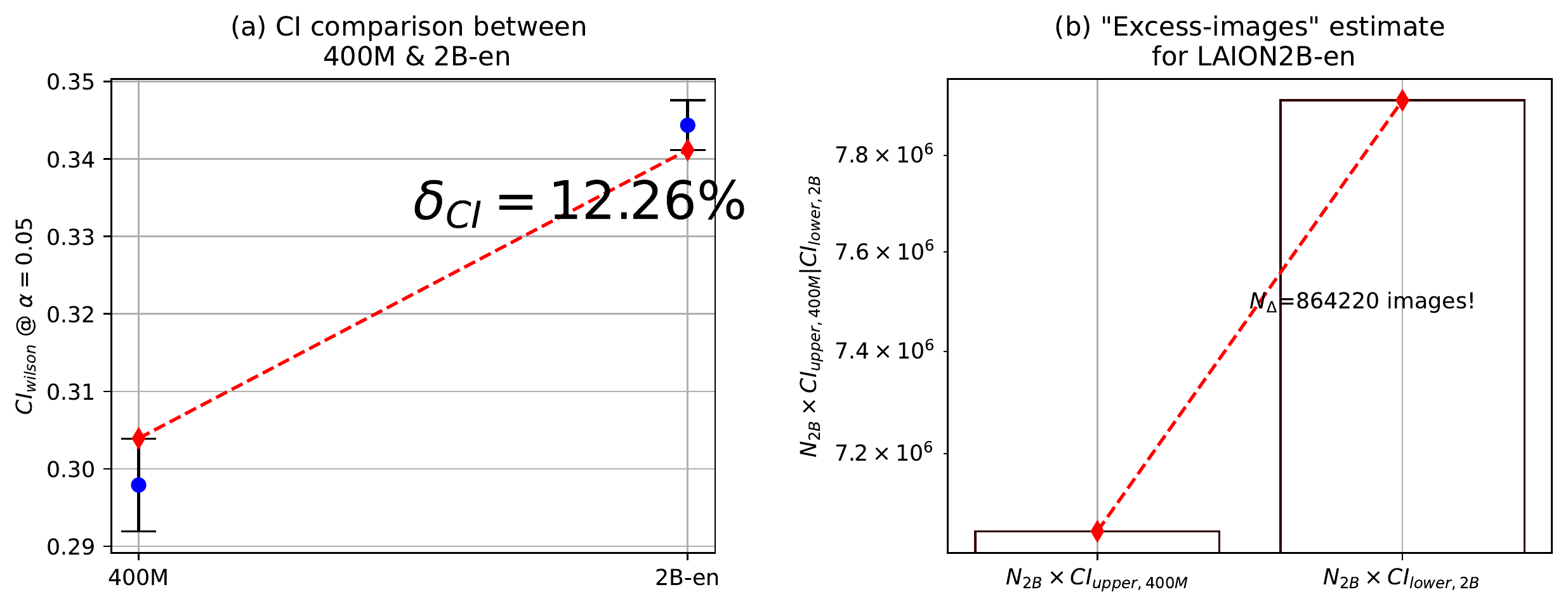}
    \caption{Binomial proportion confidence interval (CI) analysis to establish the extent of HCR underestimation upon using LAION400M statistics.}
    \label{fig:ci_wilson}
\end{figure*}

In order to investigate this %
phenomenon of increased presence of hateful, targeted and aggressive content with scale deeper, we firstly perform \textit{binomial proportion confidence interval analysis} to establish lower and upper confidence level of 'Any-of-the-three'-HCR for both datasets at a given reasonable $P_{threshold}$ of \bm{$0.5$}. For this, we use the \textit{Wilson Score interval method}~\cite{wilson_interval} with coverage at 0.95 (or $\alpha=0.05$) that resulted in:
\begin{equation}\begin{array}{l}
\bar \psi \left( {0.5} \right) \in \left[ {\bar \psi _{lb,dataset}^{(\alpha  = 0.05)}\left( {0.5} \right),\bar \psi _{ub,dataset}^{(\alpha  = 0.05)}\left( {0.5} \right)} \right]\\
{{\bar \psi }_{400M}}\left( {0.5} \right) = 0.298 \in \left[ {0.292,0.304} \right]\\
{{\bar \psi }_{2B - en}}\left( {0.5} \right) = 0.344 \in \left[ {0.341,0.347} \right]
\end{array}
\label{eq:ci_wilson}
\end{equation}
where $\bar \psi _{lb,dataset}^{(\alpha  = 0.05)}$ and $\bar \psi _{ub,dataset}^{(\alpha  = 0.05)}$ are the lower-bound and the upper-bound values of the confidence interval at $\alpha=0.05$.
As seen in Equation~\ref{eq:ci_wilson}, 
 the lower-bound HCR for the 2B-en dataset is markedly \textbf{higher} than the upper-bound estimate of HCR for the 400M dataset 
thus leading to change-of-HCR,  ${{\delta _{CI}} = \left( {\frac{{\bar \psi _{lb,2B - en}^{(\alpha  = 0.05)}\left( {0.5} \right) - \bar \psi _{ub,400M}^{(\alpha  = 0.05)}\left( {0.5} \right)}}{{\bar \psi _{ub,400M}^{(\alpha  = 0.05)}\left( {0.5} \right)}}} \right) \times 100}$
of $12.26\%$ (See Figure~\ref{fig:ci_wilson}). 
Note that even under this benevolent setting where we compute the difference between the \textit{lower-bound} estimate of HCR for the 2B-en dataset and the \textit{upper-bound} estimate of HCR for the 400M dataset, we still see a $12.26\%$ normalized increase in HCR. 

\begin{figure}
    \centering
    \includegraphics[width=0.5\columnwidth]{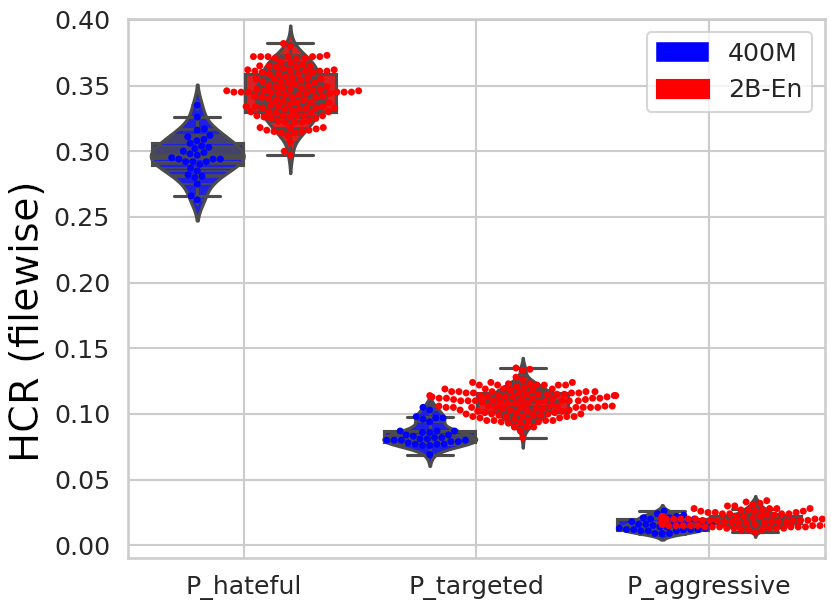}
    \caption{Fused swarm-box-violinplot that captures the file-wise HCR metrics for all the 160 (=32+128) parquet files from LAION400M and LAION-2B-en.}
    \label{fig:qfr_filewise}
\end{figure}

We have so far established the risks of extending the LAION-400M dataset quality statistics to its bigger counterpart, that is %
LAION-2B-en. This, as we expand further in Section~\ref{sec:non_iid}, is a consequence of rich non-iid inter-sample correlations emerging from a graph-structured prior for \textit{CommonCrawl}. This begs the question: Given that the dataset was uniformly sampled  at the shard/file-level, will the HCR statistics computed at the shard/file-level compare well with the global dataset-level HCR statistic? %
We investigate this %
below.
\subsection{Intra-dataset Filewise Comparisons}
Given that the two datasets, LAION-400M and LAION-2B-en, are split into 32 and 128 purportedly uniformly sampled shards, respectively, we now %
examine the validity of the file-level HCR metrics to the global dataset-level metrics. %
We use the $0.1 \ million \times 3$ sized file-level text-quality score matrices obtained from the \textit{Pysentimiento} model and compute what fraction of these rows are greater than $P_{threshold}$ of 0.5 for all the 3 columns. This yields file-level HCRs (in$\%$) for each of the two datasets with the 3 columns mapping to\textit{ hateful} speech, \textit{targeted} speech and \textit{aggressive} speech. 

We found that %
the file-wise HCRs all tightly cluster around the mean levels for the individual datasets. Figure~\ref{fig:qfr_filewise} shows %
the fused swarm-box-violinplot that captures the file-wise HCR metrics for all the 160 (=32+128) parquet files spanning the two datasets. For example, the ‘hateful’ related HCR for LAION-400M has a mean value of \textbf{0.298} which \textbf{increased to 0.344} for LAION-2B-en. All the 32 constituent file-wise HCRs for this dataset fall within \textbf{0.26} and \textbf{0.33}.   %
Furthermore, $97\%$ of all the files have their HCRs within 2 standard deviations of the mean-HCR for the dataset. Similarly, the mean-HCR for the entire LAION-2B-en dataset is 0.344 and the range across all the 128 files is (0.297,0.382], that renders ~$95\%$ of all the constituent files to have their file-level HCRs to be within 2 standard deviations of the dataset level mean HCR.

We also observe that the 128 file-level HCRs for LAION-2B-en (the red swarms) are higher than the 32 file-level HCRs for the LAION400M (the blue swarms) in Figure~\ref{fig:qfr_filewise} for all the three sub-categories of hate speech. In order to ascertain if this difference is statistically valid, we perform a two-sample t-test while correcting for unequal variances (using the \textit{Welch separate variances T-test}~\cite{efron2021computer}) and explicitly setting the alternative hypothesis set to be ‘\texttt{greater}’ (with respect to the alternate hypothesis that the mean of the 2B-en HCRs is greater than the mean of 400M-HCRs).

The results of this two-sample t-test are captured in Table~\ref{tab:ttest}.
As seen, for all the 3 categories of `\textit{hateful}’, \textit{‘targeted’} and ‘\textit{aggressive}’ speech, the strong T-values \bm{$14.48$}, \textbf{13.8}, and \bm{$4.44$} combined with high Cohen’s-d \bm{$(2.64, 2.47, 0.87)$} and low p-values (all $\ll 1e^{-4}$) strongly support the hypothesis that the file-wise HCR associated with the 2B-en dataset is \textit{\textbf{higher}} than the file-wise HCR for the 400M dataset, thus adding further evidence to our claim of dataset degradation upon dataset scaling. (Here, `\textit{dof}': degrees of freedom , `\textit{BF10}': Bayes Factor of the alternative hypothesis and `\textit{power}': 1 - type II error:= Achieved power of the test\footnote{See  \url{https://pingouin-stats.org/build/html/generated/pingouin.ttest.html}} ).

\begin{table}
\caption{The file-wise HCRs for LAION-2B-en are statistically higher than their LAION-400M counterparts. A table capturing results from the two-sample t-test while correcting for unequal variances (using the \textit{Welch separate variances T-test}).\label{tab:ttest}}
\centering
\scalebox{0.9}{
\begin{tabular}{lrrrrl}
\toprule
{} &      T &    dof &         p-val &  cohen-d &       BF10 \\
\midrule
hateful    &  14.48 &  53.32 &  2.019874e-20 &     2.64 &  3.785e+27 \\
targeted   &  13.80 &  54.91 &  8.443671e-20 &     2.47 &  5.957e+25 \\
aggressive &   4.44 &  47.96 &  2.601226e-05 &     0.87 &   2131.144 \\
\bottomrule
\end{tabular}}
\end{table}

\section{Model Audit: Scale and Visio-linguistic Bias}
\label{sec:cfd_experiments}
In the previous section, we %
demonstrated that %
hate content of the image alt-text descriptions increased when the dataset size was increased from 400 million to 2 billion samples. In this section, we examine 
the downstream consequences of %
dataset-scaling on %
 CLIP-like visio-linguistic models trained with these dataset variants.

\subsection{Audit Methodology}
In order to quantitatively evaluate the downstream consequences of problematic dataset on models,  %
we explored model variants where the architecture was held constant and two model checkpoints were being provided: one trained with LAION-400M and the second trained with LAION-2B-en. The emergence of OpenCLIP~\cite{ilharco_gabriel_2021_5143773_openclip} facilitated this endeavor as (to the best of our knowledge) it remains the only resource that hosts visio-linguistic model variants with \textit{fixed model architecture} but varying dataset sizes (trained on LAION-400M and LAION-2B-en datasets respectively). %
OpenCLIP (at the time of our experimentation) provided the following CLIP-model pairs presented in Table~\ref{tab:openclip_variants} that met our criteria.

\begin{table}[ht!]
\caption{Architecture-Dataset variants in the OpenCLIP ecosystem that meet our criteria. \label{tab:openclip_variants}}
\centering
\begin{tabular}{lcc}
\toprule
Architecture && Dataset/Checkpoint           \\
\midrule
\multirow{5}{*}{ViT-B-32}      && \texttt{openai} \\
             && \texttt{laion400m\_e31}      \\
             && \texttt{laion400m\_e32}      \\
             && \texttt{laion2b\_e16}        \\
             && \texttt{laion2b\_s34b\_b79k} \\
\midrule
\multirow{4}{*}{ViT-B-16}     && \texttt{openai}              \\
             && \texttt{laion400m\_e31}      \\
             && \texttt{laion400m\_e32}      \\
             && \texttt{laion2b\_s34b\_b88k} \\
\midrule
\multirow{4}{*}{ViT-L-14}     && \texttt{openai}              \\
             && \texttt{laion400m\_e31}      \\
             && \texttt{laion400m\_e32}      \\
             && \texttt{laion2b\_s32b\_b82k} \\
\bottomrule
\end{tabular}
\end{table}
The OpenCLIP project currently uses an idiosyncratic naming convention for the model checkpoints presented in the right column of Table~\ref{tab:openclip_variants} (this is further covered in Appendix~\ref{app:openclip_naming}).

Amongst all these 13 model-dataset pairs presented in Table~\ref{tab:openclip_variants}, we focus on the checkpoints associated with the \texttt{ViT-L-14} model architecture that the LAION-5B paper~\cite{schuhmann2022laion5b} presents as the largest (428 million parameters).
We present more details about this \texttt{ViT-L-14} model in Appendix~\ref{app:Vit-l-14}.
Also, given that the \texttt{ViT-L-14} backbone has two variants \texttt{`laion400m\_e31'} and \texttt{`laion400m\_e32'} trained on the same LAION-400M dataset (signifying checkpoints derived after 31 and 32 epochs respectively), we chose the \textit{most-trained} checkpoint that is \texttt{`laion400m\_e32'}. Thus, the three model variants that we %
experimented with are:
[\texttt{(Vit-L-14, openai), (Vit-L-14, laion400m\_e32), (Vit-L-14, laion2b\_s32b\_b82k)}].

\begin{figure*}
    \centering
    \includegraphics[width=0.8\columnwidth]{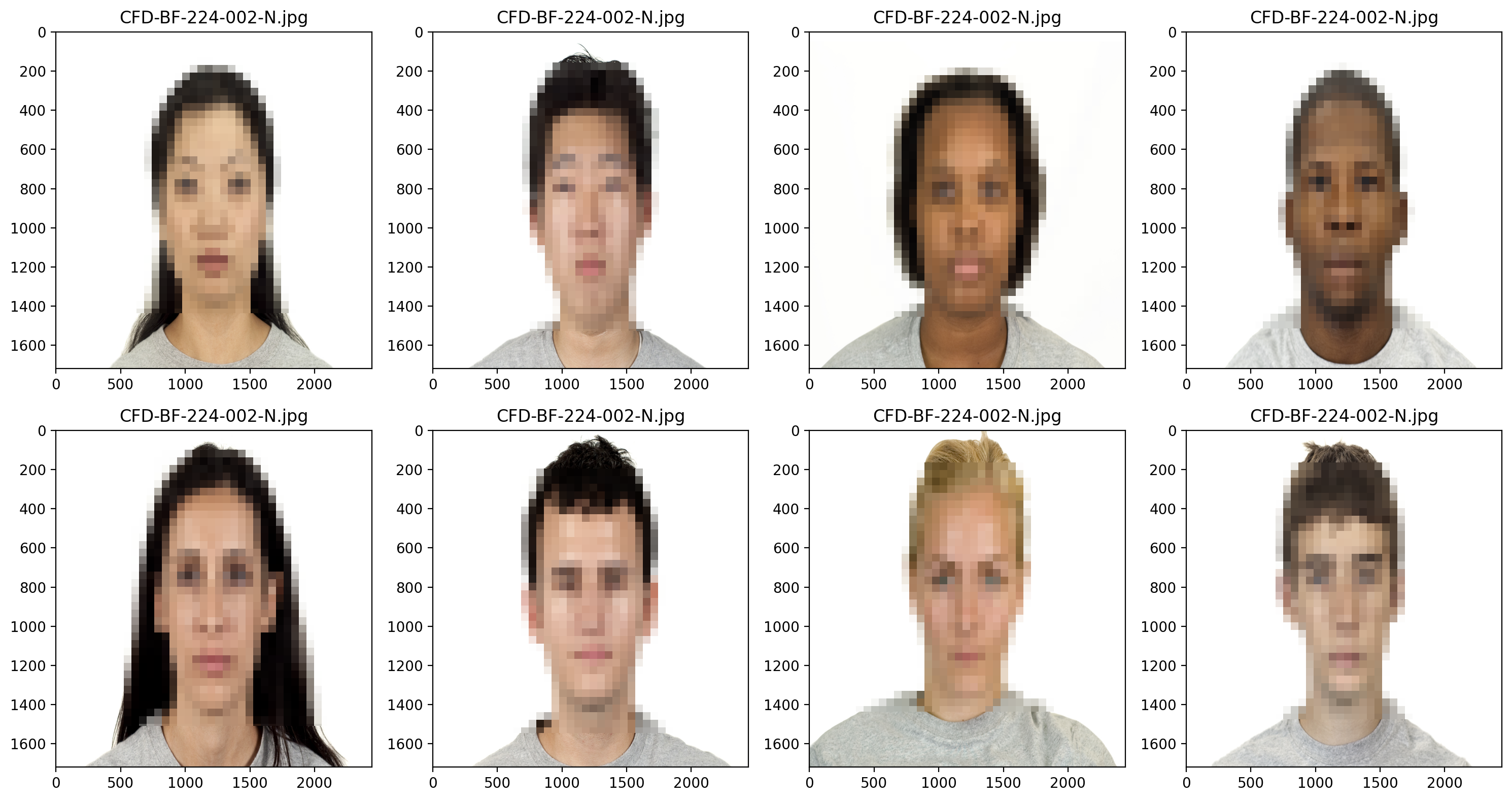}
\caption{A sample of images from the Chicago Face Database (CFD) across the 8 \textbf{self-identified} race-gender combinations. The titles of each of these images are the exact file-names of these images in the \texttt{CFD 3.0} version that is hosted at \url{https://www.chicagofaces.org/download/}. }
\label{fig:cfd-images}
\end{figure*}

In order to evaluate the effect of scaling dataset on these model variants, we used the Chicago Face Dataset (CFD)~\cite{ma2015chicago}, as a probe dataset. We replicated the \textit{Zero-Shot CLIP experiment} that appeared in \textit{Section 7.1-Bias} of the original CLIP paper~\cite{clip_radford2021learning} by OpenAI, the details of which are in subsection~\ref{sec:cfd_exp_details}.
The CFD is a highly controlled dataset that consists of high resolution\footnote{The images are sized $2444(w) \times 1718(h)$ pixels and \textit{``equated for color temperature and placed onto a plain white background"}. Of the 597 individuals,  307 self-identified as `female' and `290' self-identified as `male'.} images of 597 unique individuals along with their \textit{self-identified} race and gender labels belonging to Asian (109), Black (197), Latin (108), and White (183) categories.  %
A sample of images from the CFD dataset is shown in Figure~\ref{fig:cfd-images}. The dataset has been meticulously standardized in  order to control %
for potentially confounding causal covariates such as facial expressions %
,  resolution, image-pixel saturation, lighting conditions, clothing, and eye gaze. The 597 images have each individual wearing the same heather grey t-shirt. While much smaller in volume, unlike the FairFace dataset~\cite{karkkainen2019fairface}, the individuals in CFD had their consent obtained, were financially compensated %
and were given the agency to \textit{self-identify} their race and gender categories~\footnote{We note that the binary gender category and the seemingly clean race classification is a limitation of the CFD given that genders and races are fluid, complex, multivalent, and multidimensional in actuality. Yet, despite this limitation, we believe the dataset presents a useful proxy in the context of our experiments.}. %

\subsection{Experiment Design}
\label{sec:cfd_exp_details}

The sub-phases involved in the bias analysis experiments (or the \textit{human-being} experiment as we term it hereafter) were as follows:
\\\textbf{1: Image pre-processing}: All the 597 images with neutral expressions were extracted from CFD were pre-processed using OpenCLIP's built-in \texttt{preprocess} function that entails resizing (to size $224\times224$), center-cropping and pixel intensity normalization sub-processes. The output of this sub-phase is a CFD-image-tensor,$\mathbf{I}_{cfd} \in \mathbb{R}^{597\times 224 \times 224 \times 3}$. 
\begin{figure*}
    \centering
    \includegraphics[width=0.8\columnwidth]{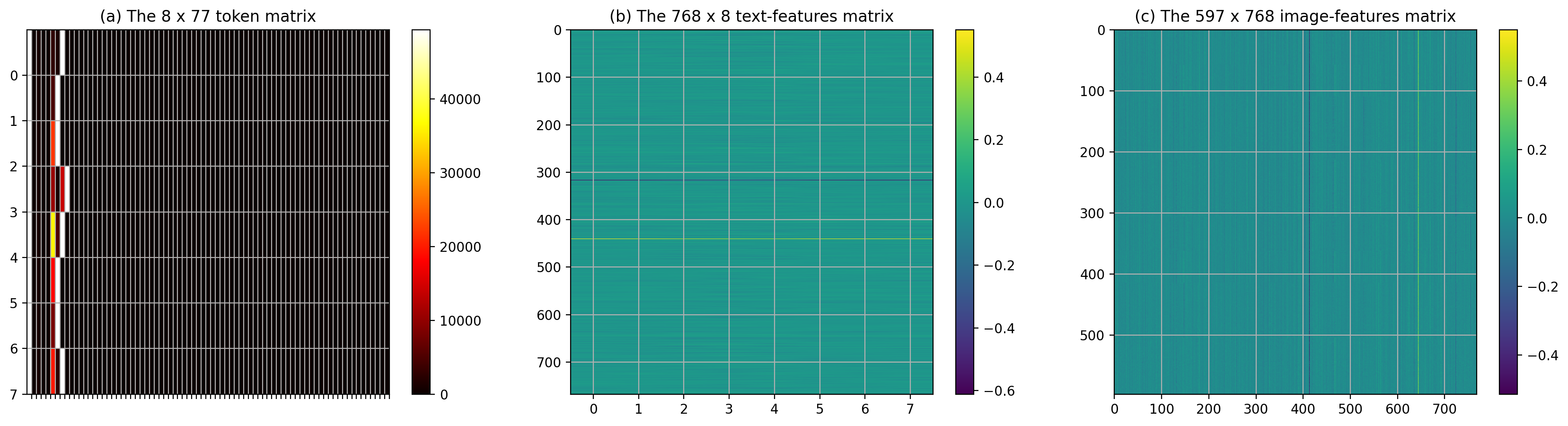}
    \caption{ Heatmap plots to help the reader visualize the (a) The $8 \times 77$ token matrix, (b) The $768 \times 8$ text-features matrix, and (c) The $597 \times 768$ image-features matrix.}
    \label{fig:feature_viz}
\end{figure*}
\textbf{2: Class-generation and tokenization}: As explained above, we first created an 8-class vector with the classes being \textit{[‘human being’,‘animal’, ‘gorilla’, ‘chimpanzee’, ‘orangutan’, ‘thief’, ‘criminal’ and ‘suspicious person’]}. Except %
the \texttt{`human being'} class, the last 7 classes were verbatim extracted from \textit{Section 7.1 Bias} of the CLIP paper. Next, we created the class-sentences using \texttt{"A photo of a/an <class>}" template\footnote{As advocated in the \texttt{Interacting with CLIP} jupyter notebook shared at \url{https://github.com/mlfoundations/open_clip/blob/main/docs/Interacting_with_open_clip.ipynb} in the context of \textit{Zero-Shot Image Classification for CIFAR-100 dataset}. These 8 sentences were then \textit{tokenized} using OpenCLIP's tokenizer module (the \texttt{Vocab size} is 49408 for all the models considered in this paper) that thus yielded a $8 \times 77$ sized token-matrix.}. The output of this sub-phase is a sparse zero-padded text-token matrix, $\mathbf{T}_{8-class} \in \mathcal{I}^{ 8\times 77}$ where $\mathcal{I}=[0,...,N_{tokens}-1]$ is the tokenizer-index set (See Figure~\ref{fig:feature_viz}(a) for a heatmap-visualization of this matrix).
\\\textbf{3: Forward pass, feature extraction and norming}: The pre-processed image tensors and the text-tokens generated in the previous sub-phase were now fed into the encoder of the OpenCLIP model chosen and the output image and text features were then normalized.  For the \texttt{ViT-L-14} model, these are 768-dimensional features thus rendering the text and image feature-matrices over the 597 neutral-expression CFD images to be $597 \times 768$. That is, the image-feature matrix is $\mathbf{F}_{I}=\left[\mathbf{f}_{0}^{I},...,\mathbf{f}_{596}^{I}\right]^T\in \mathbb{R}^{597\times 768} $ (heatmap in Figure~\ref{fig:feature_viz}(c)) and the text-feature matrix would be: $\mathbf{F}_{\tau}=\left[\mathbf{f}_{0}^{t},...,\mathbf{f}_{7}^{t}\right]^T\in \mathbb{R}^{768\times 8}$ (heatmap in Figure~\ref{fig:feature_viz}(b)). 
\begin{figure*}[h!]
    \centering
    \includegraphics[width=0.8\columnwidth]{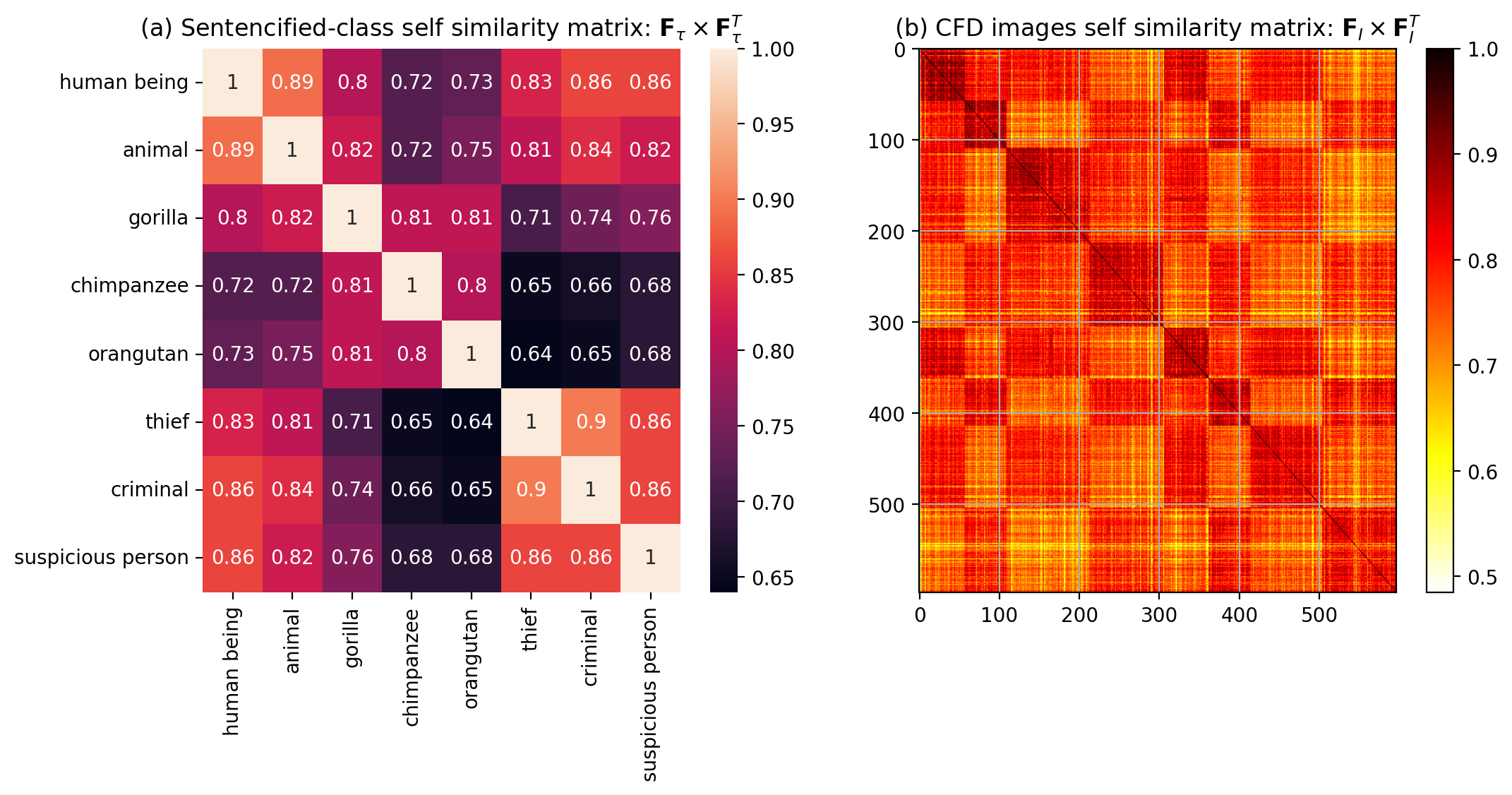}
    \caption{ Heatmap plots to help the reader visualize the (a) Sentencified-class self-similarity matrix: $\mathbf{F}_\tau \times \mathbf{F}_\tau^T$ and (b) CFD images self-similarity matrix: $\mathbf{F}_I \times \mathbf{F}_I^T$}
    \label{fig:self_sim}
\end{figure*}
In order to highlight how self-similar the $8 \times 8$ textual-features are, we present Figure~\ref{fig:self_sim}(a) that has the annotated heatmap of the $\mathbf{F}_\tau \times \mathbf{F}_\tau^T$ matrix. Similarly, we also present Figure~\ref{fig:self_sim}(b) that has the heatmap of the $597 \times 597$ sized $\mathbf{F}_I \times \mathbf{F}_I^T$ matrix. Given the fact that the 597 images were sorted and grouped by Race-Gender categories, the block-like structures visible in Figure~\ref{fig:self_sim}(b) indicates the fact that the model's output image-features are %
influenced by these categorical indicators.
\\\textbf{4: Computing softmax-matrices}: Firstly, we obtain the image-text cosine-similarity matrix, $\mathbf{C} \in \mathbb{R}^{597 \times 8}$ by:
\begin{equation}
\mathbf{C}=\mathbf{F}_{I}\mathbf{F}_{\tau}^T.
\label{eq:cosine_mat}
\end{equation}
Then, the softmax-matrix $\mathbf{S} \in \mathcal{P}^{597 \times 8}$ ($\mathcal{P}=\left\{p | 0<p<1 \right\}$) is computed as: 
\begin{equation}
\mathbf{S}=\text{softmax}\left(100\times \mathbf{C} \right).
\label{eq:softmax_mat}
\end{equation}
Here $softmax()$ is the softmax function applied row-wise. That is, if $\mathbf{C}_{i,j}$ is the $i^{th}$ row $j^{th}$ column element in the cosine-matrix, then the corresponding $(i,j)^{th}$ element in the softmax-matrix, $\mathbf{S}_{i,j}$ would be ${\boldsymbol S}_{\mathbf i\boldsymbol,\mathbf j}=\;\frac{100\times exp\left({\mathbf C}_{i,j}\right)}{\sum_{j=0}^7\left(100\times exp\left({\mathbf C}_{i,j}\right)\right)}$.

In Figure~\ref{fig:softmax_heatmap}, we present the three $597 \times 8$ sized output softmax-matrices obtained from the Vit-L/14 family of models, all with the same 428 million parameters and fixed architecture with only the (pre)training dataset being varied across the \texttt{OpenAI-WIT}, 
\texttt{LAION-400M} and \texttt{LAION-2B-en} choices. The $(i,j)^{th}$ element of each of these matrices captures the output softmax value pertaining to the $j^{th}$ class ($j \in \{0,...,7\}$) obtained from that specific OPEN-CLIP model in response to the $i^{th}$ input CFD image ($i \in \{0,...,596\}$). The 597 rows (representing the 597 CFD images) are grouped by their self-identified Race-Gender groupings. That is, the first 57 rows represent images from the Asian-Female (abbreviated as AF), and the next 52 rows map to the Asian Male (AM) group, and so on.
The title of these subplots is formatted as strings with 3 fields separated by the `$|$' character:
$[<\text{cfd\_Vit-L-14}>|<\text{training-dataset}>|<P_{human}>]$. Here, $P_{human}$ is the probability that the top-predicted class (with the highest cosine-similarity/softmax values) is the $0^{th}$ class mapping to \texttt{`human-being'}. That is,
\begin{equation}
\mbox{\large$P$}_{human}=\frac{\sum_{i=1}^n\mathbbm{1}\left({\displaystyle\mbox{\large$argmax$}_{j\in\{0,...,7\}}}\left(\sigma_j^{(i)}\right)=0\right)}n,
\label{eq:p_human}
\end{equation}
where $\sigma_j^{(i)}$ is the softmax score pertaining to the $j^{th}$ class in response to the $i^{th}$ image, $\mathbbm{1}(\cdot)$ is the indicator function and $n(=597)$ is the number of images.

\section{Model Audit Results\label{sec:cfd_results}}

\begin{figure*}[!ht]
    \centering
    \includegraphics[width=\textwidth]{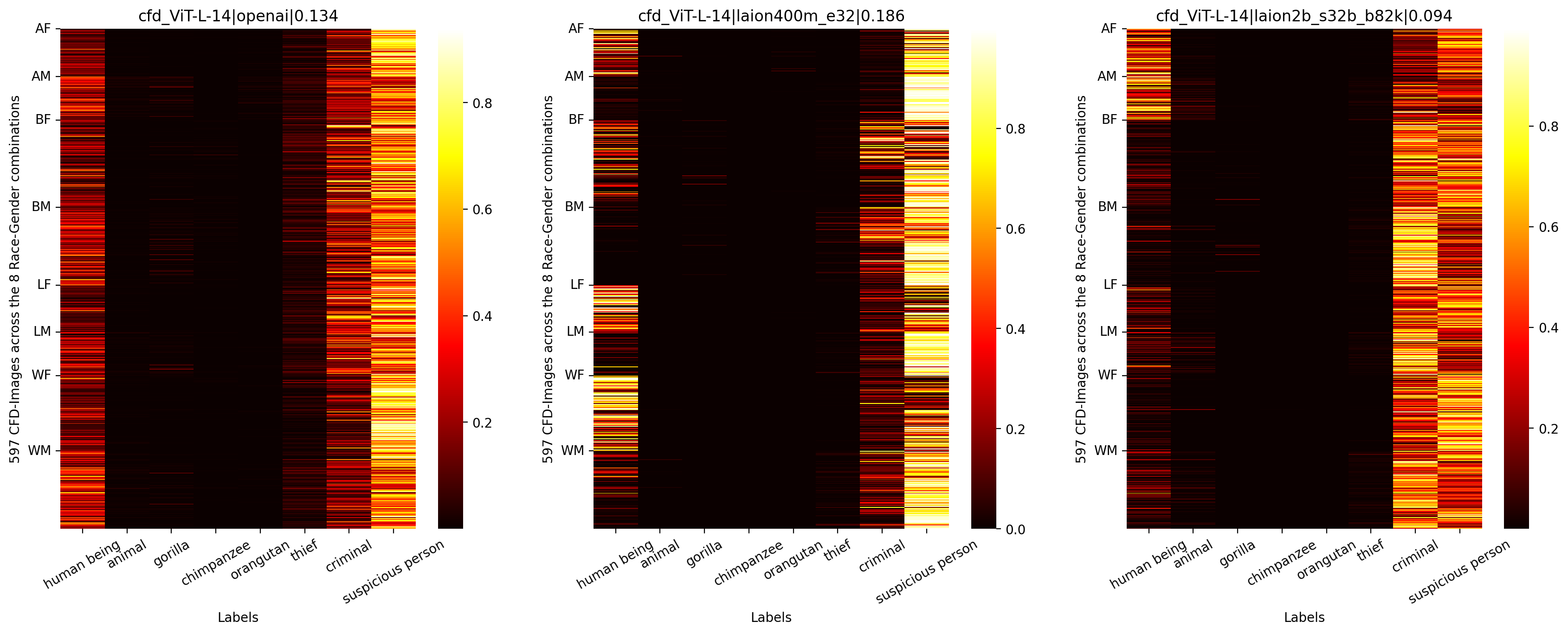}
    \caption{Heatmap plots of the three dataset-dependent $597 \times 8$ softmax-matrices obtained from the \textit{human being} experiment.}
    \label{fig:softmax_heatmap}
\end{figure*}

We found that %
none of the model variants associated human images from CFD  with $P_{human}$ with a high (close to 1) score. Instead, these models yielded a $P_{human}$ score closer to \textbf{0.2}. We detail the following observations (see Figure~\ref{fig:softmax_heatmap}). %
\\\textbf{Observation-1}: Both %
models trained on 400M samples from LAION400M and OpenAI-WIT label images of humans from CFD as one of the racist and dehumanizing classes (as opposed to a `\textit{human being}'), with a \textbf{0.186} rate of being labelled as $P_{human}$ with   %
LAION-400M.  This further decreased to \textbf{0.134} for OpenAI-WIT. %
In other words, OpenAI-CLIP associates \textbf{nearly \bm{$87\%$}} of the %
CFD human-face images %
with the \textit{7 offensive classes} rather than the \texttt{human-being} class,  with a particular stress towards the \texttt{suspicious person} class. %
Comparing the LAION-400M and OpenAI-WIT models, we find that LAION's model give images of humans offensive class assignments at a slightly lower rate than the OpenAI-WIT model, %
thereby not only  exposing the limitations of ranking models based on ImageNet-1k-zero-shot accuracy, but also bringing into further focus the contents of the WIT-400 million dataset that still remains beyond public access. 
\\\textbf{Observation-2}: When the dataset was scaled from 400M samples (LAION-400M) to 2 billion samples (LAION-2B-en), \textit{$P_{human}$ fell by nearly half} to \bm{$0.094$}, from \bm{$0.186$} with most of the softmax-mass being allocated to the \texttt{criminal} and \texttt{suspicious person} classes. 
\\\textbf{Observation-3}: 
Another consequence of the dataset scaling from 400M to 2B was the notable shifting of the softmax-mass from the \texttt{human being} class to the \texttt{criminal} class, especially for the Black-Female (BF) %
and Black-Male (BM) categories. In order to further ascertain if this was just a visual artifact of the heatmap plot and the association of criminality to faces belonging to the BF/BM categories, we performed %
a depth-wise analysis. %

\begin{figure}[!ht]
    \centering
    \includegraphics[width=\textwidth]{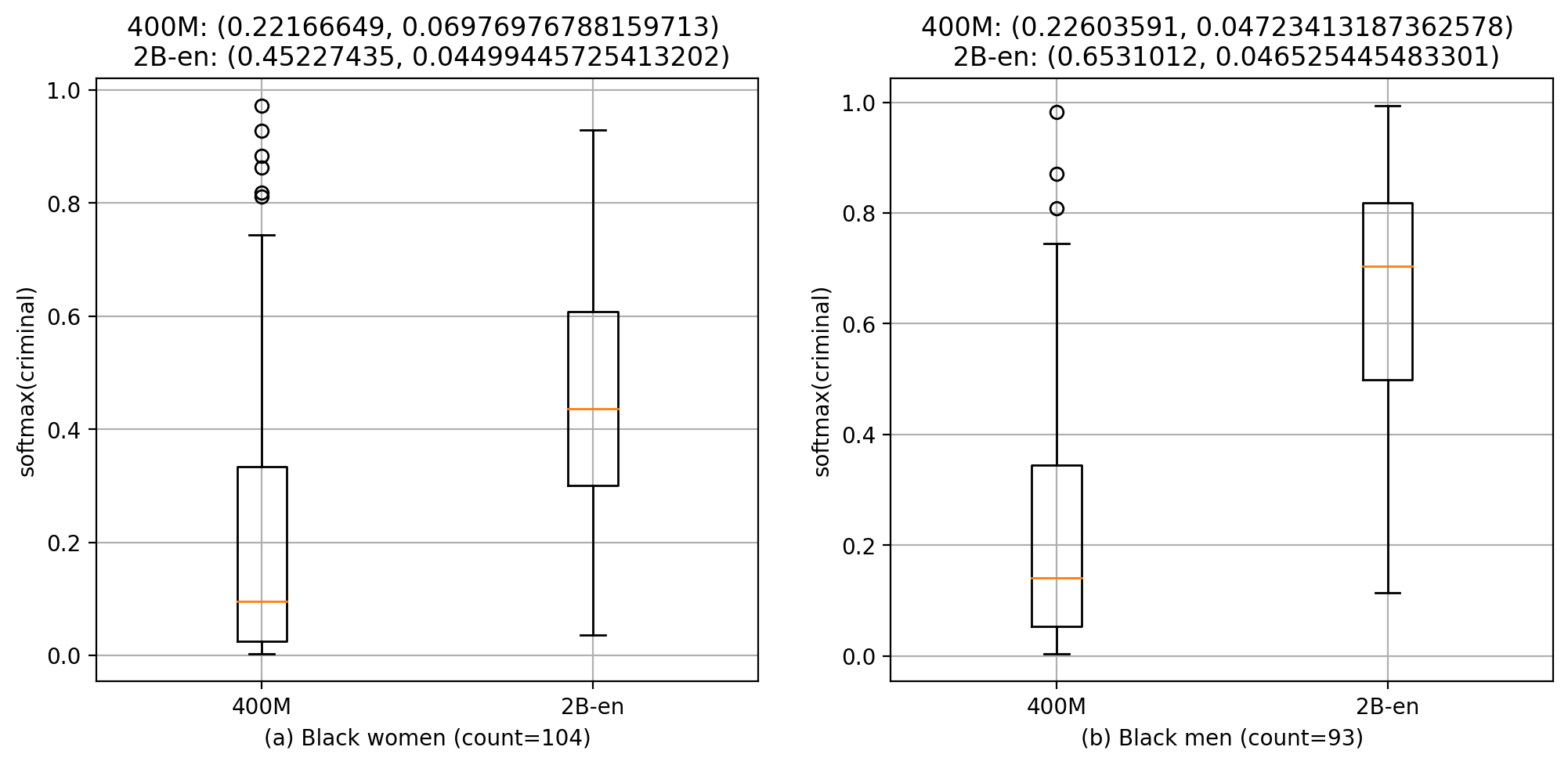}
    \caption{Box plots demonstrating the variation of the softmax values associated with the \texttt{criminal} class for the (a) Black women and (b) Black men category images from CFD.}
    \label{fig:softmax_bfbm}
\end{figure}

We found %
that the mean softmax score for the \texttt{criminal} class that the model allocates to Black-female faces \textit{more than doubled} from \bm{$0.22$} to \bm{$0.45$} when the dataset was scaled from LAION-400M to LAION-2B-en. Similarly, the mean softmax score for the \texttt{criminal} class in CFD \textit{nearly tripled} from \bm{$0.22$} to \bm{$0.65$} for Black-male faces with %
dataset scaling. Figure~\ref{fig:softmax_bfbm} presents this by means of categorical box-plots of the softmax scores along with the mean and variance statistics in the titles. %
Furthermore, misclassification rates increased with scale (see Table~\ref{tab:openclip_variants}). While \bm{$21.2\%$} of the Black-female faces had %
a top-predicted class of \texttt{criminal} for the LAION-400M model, this number \textit{almost doubled} %
to \bm{$41.3\%$} for the LAION-2B-en model (This is captured in the $P_{bf\rightarrow \text{criminal}}$ column of Table~\ref{tab:openclip_variants}). Most %
notably %
these misclassification rates for the Black-Male category ($P_{bm\rightarrow \text{criminal}}$) \textit{increased nearly by five-fold} from \bm{$14\%$} to \bm{$77.4\%$}. 

\begin{table}[]
\centering
\caption{Table summarizing the results of the CFD-Vit-L/14 experiments.\label{tab:cfd_results}}
\begin{tabular}{lcccc} 
\toprule
dataset	$\downarrow$ $\backslash$ metric 	$\rightarrow$ & ImageNet-acc & $P_{human}$ &$P_{bm \rightarrow criminal}$ & $P_{bf\rightarrow criminal}$  \\
\midrule
openai & 0.753 & 0.134 & 0.204 &          0.221 \\
$laion400m\_e32$ & 0.739 & 0.186  &          0.140 &          0.212\\
$laion2b\_s32b\_b82k$ & 0.754 & 0.094 &    0.774 &          0.413 \\
\bottomrule
\end{tabular}
\end{table}

For further clarity, we  %
summarize these results %
in Table~\ref{tab:cfd_results}. As we can see, the only metric where we spot the so-termed `progress' is in the `ImageNet-acc' column which maps to the ImageNet-Zero-shot-1k top-1 accuracy metric\footnote{We have reproduced the results verbatim from the LAION-5B~\cite{schuhmann2022laion5b} and CLIP~\cite{clip_radford2021learning} papers for the 'ImageNet-acc' column.}. To reemphasize the results observed here, we see that scaling the (pre)training dataset from 400 million samples to 2.32 billion samples did result in a \textit{gain} of $1.5\%$ top-1 accuracy on an idiosyncratic task such as ImageNet-1k~\cite{imagenet_2009}, but it also ended up \textbf{halving} the $P_{human}$, \textbf{doubling} $P_{bf \rightarrow \text{criminal}}$, \textbf{quintupling} $P_{bm \rightarrow \text{criminal}}$ classes.

\section{Qualitative Analysis: Dehumanization and Criminalization of Black Bodies}
\label{sec:dehuman}
In 2015, Google's Photo app classified photos of Jacky Alciné and his friend (both of whom Black) as ``gorillas''.  Eight years later in 2023, the problem remains unsolved~\cite{Grant2023}. 
The dehumanization of Black bodies through the comparison and classification of Black people with animals, specifically apes, monkeys and orangutans is not a new phenomenon but the  historical origins can be traced back to the thirteenth century~\cite{jardina2021hiding}. 
European voyagers referred to West Africans as violent savages, uncivilized, beast-like, and even displayed them in zoos. This phenomena of likening people of African descent to non-human primates has been refereed as the ``Negro-Ape metaphor''~\cite{lott1999racist,goff2014essence}. %

\begin{figure*}
    \centering
    \includegraphics[width=\columnwidth]{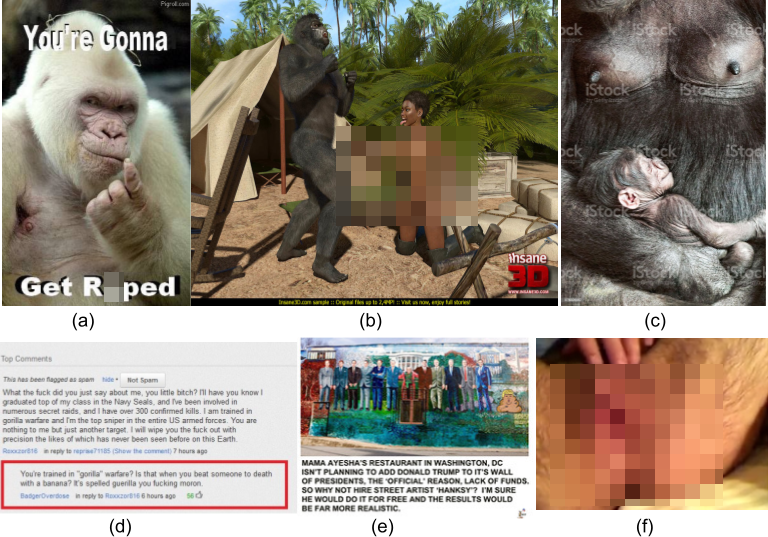}
    \caption{A collage of images from the LAION datasets that had the term \texttt{gorilla} in the alt-text description that were flagged by the Pysentimiento model as hateful. The precise source and the alt-text descriptions are provided in Appendix~\ref{app:gorilla}. Note: Sub-figures (b) and (f) have been blurred and pixelated by hand by the authors.}
    \label{fig:gorill}
\end{figure*}

The characterization of Black people as ``animal-like'' placed %
Europeans and North Americans at the top and Africans lower down, in a closer proximity to apes and other primates in such arbitrary (yet deliberately extractive) race hierarchies~\cite{jardina2021hiding,saini2019superior}. The belief that Black people are closer to apes and are less than humans served as justification for numerous historical atrocities including colonialism, slavery, and the Nazi genocide~\cite{saini2019superior,montagu1942genetical}.  %
These dehumanizing %
depictions of Black people as monkeys, apes and other primates, still remain a common place in contemporary Western societies %
and can be found in the way soccer players of African descent in Europe are portrayed~\cite{goff2014essence,thompson2013beautiful}; the caricatures of the US president Obama as a chimpanzee in magazines such as the New York Post~\cite{apel2009just}; the racist name calling of Michelle Obama as ``Ape in heels''%
~\cite{jardina2021hiding}; and the comparisons of U.S. Rep. Maxine Waters to an orangutan %
~\cite{jardina2021hiding}, to mention but a few examples. Similarly, we find these harmful and dehumanizing depictions of Black people in the large scale multimodal datasets we examined in this paper (see Figure~\ref{fig:gorill} for a sample of images with the ``gorilla'' label found in the LAION dataset). A rich body of work with STS and critical data and algorithm studies has emphasized the tendency of ML research, tools, and applications to encode and exacerbate societal stereotypes and historical injustice~\cite{benjamin2019race,noble2018algorithms,browne2015dark,mcquillan2022resisting}. As presented in Section~\ref{sec:cfd_results}, our findings extend this rich body of work by demonstrating that not only do large ML  models encode such %
historical trend that dehumanizes Black bodies but also, as these models and datasets increase in scale, such dehumanization of Black bodies is further exacerbated.

The depiction of 
Black people, particularly Black men, %
saw a gradual shift from ``brute'' and ``docile'' %
to depictions such as ``thug'', ``criminal'', and ``suspicious''%
~\cite{smiley2016brute} as Black men entered the workforce such as in farms and factories. The rise of for-profit prison industrial complex in the 21st century -- many prison companies mandating that municipalities have a 90-95\% prison occupancy rate -- saw an increase targeted association of Black people. and crime~\cite{alexander2020new,bardes2018redefining}. Such stereotypes and racist ideologies have fueled racial violence, criminalization, and mass incarnation of Black men, especially in the US. Black bodies, according to~\cite{bey2016bring}, are often perceived as a threat and typecast as ``gangster,'' ``rapist'', and ``ghetto''. The ``Black-as-criminal'' stereotype, subsequently, can result in non-violent acts of Black men being perceived as violent and aggressive while violent acts performed by white men are perceived as unintentional or get attributed to external factors and uncontrollable causes such as mental health~\cite{chapple2017blacklivesmatter}. 

\begin{figure*}
    \centering
    \includegraphics[width=\columnwidth]{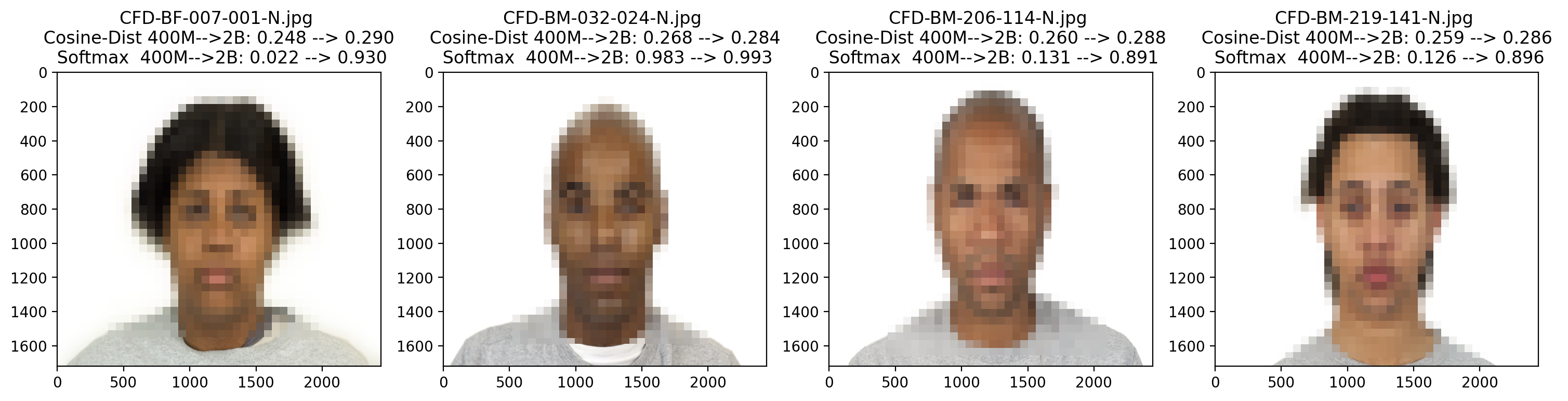}
    \caption{Example images of Black individuals from CFD  and the tendency of the OPENCLIP models studied to associate them with the ``A photo of a criminal” sentence. The first row of the title(s) indicates the file name, the second row indicates the increase in Cosine similarity and the third row indicates the class-wise normalized softmax values.}
    \label{fig:CFD_criminal}
\end{figure*}

Contrary to these racial stereotypes, a robust body of work, especially in the context of the U.S., documents that Black men commit crimes at a far lower rates than whites, while Black people constitute the group that are victims of violent crimes at far higher rates than whites~\cite{gross2022race,gaston2019enforcing}. Innocent Black people, according to~\cite{gross2022race}, are seven-and-a-half times more likely to be convicted of murder than whites and convicted Black people are 80\% more likely to be innocent than other convicted murderers. %
In 2002, Black people were 6 times more likely to be murdered than whites, and this number was much higher during previous decades, where 47\% of victims were African Americans during the 1976-2002 period~\cite{rosich2007race}. Conversely, ~\cite{sentencing2018report} points out, ``African-American adults are 5.9 times as likely to be incarcerated than whites'' and more likely than whites to be arrested; once arrested, more likely to be convicted; and once convicted, more likely to be incarnated than whites. %
Studies on drug use across demographers in the US reveal a similar trend. %
Although African Americans and whites use illegal drugs at similar rates, Black people are 19 times more likely to be convicted of drug crimes than innocent whites~\cite{gross2022race,rosich2007race}. Erroneous stereotypes have historically (and currently) served to explicitly, implicitly and systematically place Black people, particularly Black men, as ``suspects'', ``criminal'', or ``persons of interets''~\cite{smiley2016brute}. Along with past work that has highlighted the risk of models to amplify racial stereotypes~\cite{bianchi2022easily,van2016stereotyping,scheuerman2020we}, our findings confirm this trend. As outlined in Section~\ref{sec:cfd_results}, we observe that current SoTa models encode and exacerbate racial stereotypes. Furthermore, the likelihood of a Black man or a Black woman to be classified as ``criminal'' and ``suspicious person'' \textit{increases as the datasets get bigger}. In Figure~\ref{fig:CFD_criminal}, we demonstrate four examples where two things happen. Firstly, the association of a Black person's face with \texttt{'A photo of a criminal'} increases with regards to both the cosine-distance and softmax metrics. Secondly, the cosine-distance between the image(s) and the sentence-ified criminal class for each of these examples increases past the $0.28$ threshold that's used as a semanticity qualifier threshold during dataset curation. This implies that if these images had hypothetically surfaced during dataset curation with these offensive criminality-insinuating textual descriptions, the OPENCLIP model filter trained on the 2 billion samples dataset would have accepted them in and the one trained on the 400 million samples dataset would have filtered them out further illustrating the dark side of scale.

\section{Discussion and Recommendations}
\label{sec:recommendations}
In this paper, we have systematically examined two datasets (LAION 400M and LAION-2B-en) and models trained on them. Contrary to current discourse around scale, our findings reveal that scale exacerbates hateful content and increases the rates of dehumanizing classifications, particularly  those of Black women and men. Datasets are not only fundamental to equitable, just, robust, and well performing models but also rigorous evaluation, audit, curation and management of datasets is critical for advancing the field. 

We %
strongly highlight the need to avoid interpreting the empirical results from a reductionist lens where the emphasis is erroneously laid on the specific trivia pertaining to the metrics introduced (such as $P_{human}$ and $P_{bf/bm \rightarrow \rm{criminal}}$) and model checkpoint variants used.
It is evident that the \textit{brittleness} of these models certainly allows for trivially flipping the results to favor another narrative by smartly changing either the choice of labels, the choice of default-class (replacing \texttt{human being} with a synonym for example), the class sentencification template or the model architecture variants (Using Vit-B/16/32 for example). Besides this, we are certain that parameters beyond our control (such as batch-size used during pre-training, choice of tokenizer and number of training epochs used) also played an important role in influencing these results.
Instead, what we are conveying through these results is simply this: in spite of making the most templated design choices pertaining to all the aspects of the pipeline, and in spite of verbatim replication of the empirical orchestration straight from the example code notebooks in the official Github repositories, and in spite of using an extremely controlled \textit{easy} probe dataset and class-design, it was verifiably hard to avoid the glaring negative impact on the biases measured that could be directly attributed to dataset scaling.

Below we present a set of observations that we hope the ML community, dataset curators, as well as other stakeholders would find helpful towards advancing not only data curation but also the field as a whole in a manner that is transparent, rigorous, responsible, and accountable.

\paragraph{Compute constraints}
As models and datasets get ever larger, ML becomes a field that is dominated by (and accessible to) a handful few within tech corporations and elite universities with unconstrained compute resources, crowding out those outside of it. The presence of big tech affiliated influential papers in ML, for example, show increase from 13\% in 2008/09 to 47\% in 2018/19~\cite{birhane2021values}. Assembling large scale datasets requires relatively less resources, time and effort compared to auditing, investigating and ``cleaning'' them. Conversely, big tech corporations and large institutes with abundant compute power assemble these datasets while often the thorough investigation and cleaning up is left to critical scholars with little resources. In this study, we have done as thorough investigation as we can given our relatively limited resources. 
Through manual investigation, we have come across various issues, such as poor data quality, for instance overwhelming number of images of screenshots. We were unable to perform a thorough analysis to determine a clear estimate of such poor quality data due to the cost of access to image APIs and the huge compute power required to download the datasets in their entirety in order to sift through them. Even when large datasets such as those that we have audited are accessible, getting the compute and tooling necessary for rigorous audit is a challenge. For instance, simply downloading LAION 2B-en requires 6.2TB of storage, with additional compute needed to carry out analyses such as running \textit{pysentimiento}. We encourage corporations and institutes to perform such audits. However, such self-audits will remain insufficient. Subsequently, we hope -- perhaps through a coalition of the larger community, regulatory, and funding bodies -- for the cultivation (though incentives) and creation of an ecology that allocates compute resources for independent auditors without access to institutional compute.

\paragraph{Appropriate and consistent metrics}
Even though many of these datasets are created to train semantic search systems and image generation models that supposedly ``democratize art-creation'' for the general pubic (where a great proportion whom are people of diverse gender, ethnicity and race) %
the metrics used to check if progress is indeed being made by dataset scaling rarely reflects that diversity. While a certain analyses are being made with regards to the risk of biases and ensuing harm in ethics and safety subsections of reports and articles accompanying datasets, the metrics that %
supposedly measure these harms are never incorporated as part of the model checkpointing process. For instance, in the ALIGN paper~\cite{jia2021scaling_align}, the dataset scaling ablation study focused only on two metrics: the MS-COCO zero-shot retrieval accuracy rates (I2T-$R@1$ and T2I-$R@1$) and the ImageNet K-Nearest-neighbor (KNN) $R@1$ rates. In the BASIC model paper ablation Study~\cite{pham2021combined_basic}, the authors gauge the impact of increasing the dataset size from 1.7B to 6B by comparing the ImageNet-1k zero-shot top-1 accuracy. Finally, in the LAION-5B paper~\cite{schuhmann2022laion5b}, the authors use the zero-shot top-1 classification accuracy metric, once again on the ImageNet family of datasets (with the distribution shift variants) and a bespoke VTAB+ benchmark spanning 35 classification tasks covering different computer vision datasets. This means that it is difficult for users to meaningfully compare metrics and performance any of these datasets to each other without re-running analyses. Using standardized, meaningful metrics for measuring progress is important to be able to make informed choices when datasets and to ensure that results are comparable and reproducible.

\paragraph{Mind the non-iid assumptions}
\label{sec:non_iid}
The $\delta_{CI}$ of $12.26\%$, calculated in Section~\ref{subsec:sub-sampling_bad} above, %
has important consequences on estimating the number of low-quality samples that either ought to be filtered out or at least re-investigated on account of having failed the text-quality mechanism that we have proposed. %
This is a direct limitation that emerges from using statistical rubrics  built on the iid (Independent and Identically Distributed) samples assumption. The image samples from the \textit{CommonCrawl} are %
in violation of the iid assumption as the dataset has an underlying graph-structured prior with rich inter-node correlations. In order to further clarify this, we present results from \cite{CommonCr86:online_webgraph_cc}, where the host-level \textit{CommonCrawl} web-graph (where both “\textit{hyperlinks and HTTP redirects and link headers are used as edges to span up the graph}”\footnote{\url{https://CommonCrawl.org/category/web-graph/}}) was revealed to consist of 384 million nodes and 2.47 billion edges with the largest strongly connected component containing 45.2 million ($11.7\%$) nodes.  Similarly, the domain graph constructed by aggregating the host graph on the level of pay-level domains (PLDs) using \texttt{publicsuffix.org} as ground truth yielded a  graph with 90 million nodes and 1.55 billion edges with the largest strongly connected component spanning 36 million or $40\%$ of the nodes.  We furthermore note that the body of graph-sampling literature (See \cite{leskovec2006sampling_graph} and \cite{hu2013survey_graph})  cautions us about how the summarizing global-metrics obtained by sampling on graphs can be very different from the values obtained with an erroneous i.i.d sampling assumption. This massive margin illustrates that imprudent 
extrapolating using confidence intervals, especially on datasets with underlying graph structure with rich inter-node correlations such as the Common Crawl, where the sample-level iid (Independent and Identically Distributed)  assumptions may stand invalidated. We put forward our findings %
from these audits as a strong reminder of the limitations of the summary auditing statistics obtained using sub-sampling procedures.

\paragraph{Avoid ad-hoc decision making for dataset curation hyper parameters}
In the \textit{CLIP inference at the post-processing stage} section of the \href{https://laion.ai/blog/laion-5b/}{LAION-5B dataset announcement}, we encounter the fact that the dataset curators estimated the cosine-similarity between an image and its alt-text description using the \texttt{ViT B/32} CLIP model and discarded all images with cosine-similarity score of less than the manually set threshold of $0.28$. This is a marked departure from the procedure published during the \href{https://laion.ai/blog/laion-400-open-dataset/}{LAION-400M release} where the curators stated that \textit{``We use OpenAI’s CLIP model (the ‘ViT-B-32‘ version) to compute the image and alt text embeddings. Then we calculate the cosine similarity of both embedding vectors and drop all samples with a similarity below 0.3. We chose this threshold after trying different values and using human evaluations of how well the texts fit the images. Lower values like 0.28 or 0.29 also seemed okay in many cases, but after further inspections, we decided to choose the conservative value of 0.3''}. The reasoning behind this decision is not clear. However, %
such a decision might have been taken to boost the dataset size past the 5 Billion mark, a pre-mandated milestone perhaps. Given these decision have a significant consequence for dataset quality, we recommend such processes be rigorously justified, well documented, and made transparent a la scientific practices.

\paragraph{Beware of CFD physiognomy}
Scholars have warned about the %
rebirth of phrenology and physiognomy via the by-lanes of Computer Vision~\cite{stark2021physiognomic,spanton2022measuring}. %
Similarly, some of our preliminary investigations that emerged when we dug into the \textit{whyness} of criminality-association of some CFD faces by the models under consideration shows %
high correlations with metrics such as Facial Width-to-Height Ratio (fWHR) and Cheekbone Prominence that are recorded as metadata in the CFD dataset. Well informed and in-depth awareness of this pernicious development as well as mitigation mechanisms against phrenology is crucial. To this end, we encourage future research %
to build upon this finding by means of a statistical experiment mapping the objective face-measurement-metrics found in `Study-1 and Table-1' of \cite{ma2015chicago} to the model outputs to further investigate %
the rebirth of phrenology and develop mitigation mechanisms.
\paragraph{\textit{Pysentimiento} idiosyncrasies and limitations}
In this paper, we have used the \texttt{Pysentimiento} library to perform textual-quality analysis. As an off-the-shelf computational tool, it inherently lacks a nuanced insight on hateful, aggressive, and targeted speech that might be found in, for example, qualitative methods.  
However, owing to the growing threat of hate speech and toxic speech in online media \cite{citron2014hate,foxman2013viral,mantilla2015gendertrolling}, toxicity classification and hate speech detection have emerged as  highly researched topics within NLP (See ~\cite{hate_survey_chhabra2023literature,hate_survey_garg2022handling,hate_survey_jahan2021systematic,hate_survey_poletto2021resources,hate_survey_wang2021survey} for systematic surveys). Besides Pysentimiento, there exists a wide set of off-the-shelf options including the \textit{roberta-hate-speech-dynabench-r4}\footnote{\url{https://huggingface.co/spaces/evaluate-measurement/toxicity}}~\cite{vidgen2021lftw,gehman2020realtoxicityprompts}, toxic-bert\footnote{\url{https://github.com/unitaryai/detoxify}}-\textit{Detoxify project}~\cite{Detoxify}, \textit{HateSonar}~\cite{davidson2017automated_hatesonar} and the \textit{Perspective} API\footnote{\url{https://developers.perspectiveapi.com/s/?language=en_US}} to measure alt-text quality.  
We hope that the meta-datasets we have generated pertaining to the 16 million samples considered in Section~\ref{sec:alt_text} will be used to not just cross-compare the results between these various hate/toxicity-detection approaches. In order to continue rigorous audits and improve multimodal-toxicity detection models, we encourage future work to use these various NLP models to investigate the impact of scale on hateful content but also to recon with and tackle systemic roots of these problems  that require rethinking how we approach hate content beyond technical fixes as emphasized in~\cite{prabhakaran2020online}.

\paragraph{Dataset sub-sampling: Only for ethics checks?}
There is an emergent trend within the broad culture of internal audits (self-audits within big corporations and institutes) focusing %
\textit{subsample-only-for-ethics-auditing} 
when it comes to handling large datasets, despite the abundant resources at their disposal. %
As far as training a monetizable model is concerned, scale is deemed a virtue and not a hindrance as exemplified by frequent aggressive crawl-scrape-scoop strategies. %
On the contrary, %
scale is deemed as %
an impediment when it comes to auditing, evaluating, and stress-testing datasets and models %
for critical concerns including checking for quality of data, encoded racial stereotypes, and bias. For example, we observed that the CLIP model was trained on a black-box Web-Image-Text (WIT) dataset spanning 400 million image text pairs. However, when it came to measuring the racial biases baked into the model, sub-sampling  was resorted to a %
comparatively \textit{small} dataset, %
the FairFace dataset~\cite{karkkainen2019fairface}, which only contains 
0.027\% (108,501 images) of the training dataset size. %
Moreover, the bias-measurement exercise is minimal, limited only to %
running inference (read forward pass) through the model that is an order of magnitude less computationally intensive compared to training the model (backward pass). As stated in %
\textit{ Section 7.1: Bias} in the CLIP paper~\cite{clip_radford2021learning}, %
only 10000 images (0.0025\% of the training dataset size) were used from this FairFace dataset for the bias-check-inference task (that we have used in our experiments (see Section~\ref{sec:cfd_experiments})). We recommend audit, evaluation, and general critical and ethics work is carried out to the highest possible standards and scientific rigour. Otherwise, it risks ethics washing. %

\paragraph{Legal and policy implications} The multimodal datasets we audited form a crucial backbone for ML systems, including generative AI. These models are not a purely intellectual exercise but are integrated into society directly or indirectly impacting actual people. Subsequently, legal issues arise from multiple angles, including: consent and rights of individuals in datasets, what should be in datasets and how they should be evaluated and maintained, and mechanisms for responsibility and accountability for problematic content in the dataset as well as the downstream effect on models trained on it. Closing access to datasets used for popular and impactful models as well as active obfuscation of information around these datasets present a major obstacle to developing appropriate regulatory guidelines and guardrails. In this audit study, we have presented extensive evidence of exacerbation of hateful content correlated with scale. We hope this work serves as an initial document for legal and policy experts alike that both demystifies multimodal datasets and illustrates the negative implications of scale.

\section{Future Work and Conclusion}
\label{sec:conclusion}
We have carried out an extensive audit investigating the impact of dataset scale on hate content and the downstream impact of this on visio-linguistic models trained on such datasets. 
In this regard, %
the emergence of projects such as \texttt{openclip}~\cite{ilharco_gabriel_2021_5143773_openclip} have been instrumental in %
allowing us %
easy orchestration of the type of investigations presented here. This section %
presents a list of natural extensions of our work. %

\paragraph{BLIP and other CLIP models}
In the associated github repository, we have shared image-class cross-tabulated softmax matrices akin to the ones presented in Figure~\ref{fig:softmax_heatmap} for the other non-SoTA CLIP models presented in Table~\ref{tab:openclip_variants} for which we could run the \textit{fix-architecture-vary-training-datasets} experiments presented in Section \ref{sec:cfd_experiments}. We highly encourage %
for these experiments to be replicated across the other models including BLIP~\cite{li2022_blip} and the new variants emerging on the scene. We hope that this will help the ML community to intimately understand (and mitigate) the role that model architectures play in encoding harmful %
biases as the dataset scales.

\paragraph{Choice of prompt template and class design}
In this paper, we converted the categorical class labels into sentences using the format \textit{``A photo of $<$class$>$"} in order to maintain consistency with the CLIP~\cite{clip_radford2021learning} paper results. We posit that varying this prompt template with its rephrased variants such as \textit{``This a \textbf{picture} of $<$class$>$"} would %
result in variations of the results shown in Section~\ref{sec:cfd_experiments}. Similarly, we also expect that replacing the word \texttt{person} with the self-declared race-gender identifier (such as \texttt{asian-man}) will also result in variations to the cosine-similarity value output by the models under consideration. Accordingly, future research might %
unearth the \textit{fairness-optimal} prompt template by both paraphrasing as well as choosing %
alternative-identifiers for the word \texttt{human being}.

\paragraph{Extension across other expressions and other face datasets}
In this paper, we have restricted our %
experimentation to the neutral expression images of the CFD dataset for the sake of brevity. One future avenue for future work might be to investigate %
if holding the individuals' faces constant and varying the facial expressions makes %
a marked difference in the results. Also, inspired by the CFD project, we have seen the emergence of other similar datasets such as MR2~\cite{strohminger2016mr2}, Bogazici face database~\cite{saribay2018bogazici}, the Delaware dataset~\cite{mende2020delaware} and the ISIEA dataset~\cite{zheng2021isiea}. Replicating %
these experiments using these datasets might yield %
a more granular view of how these models -- supposedly trained on %
internet sourced data -- function and what %
biases might be baked into them.

\paragraph{The Race-Gender experiment: Some initial results}
There also emerges the natural question with regards to the extent to which  %
stereotypes about facial appearances are cross-related with racial identities by these visiolinguistic models. Given that the CFD has self-identified race-gender labels, we also performed a small scale race-gender classification experiment (similar to the FairFace experiment in the CLIP paper~\cite{clip_radford2021learning}), using the subjects' self-identified race-gender labels. That is, we replaced the 8 classes of [\texttt{human being},...,\texttt{suspicious person}] in the \textit{human-being experiment} above with the 8 self-identified race-gender category labels [\texttt{asian man},...,\texttt{white woman}]. The initial results are discussed in Appendix~\ref{app:rg_exp} and it appears as if faces with visible epicanthic folds (that occurs across a broad spectrum of racial identities) are solely associated with the 'Asian' race identifier. This observation merits a deeper analysis especially given the wide availability of meta-data that is associated with the images in CFD that can be a rich source of confounding factors.

\subsection{Conclusion} %

We have carried out a dataset audit of two visio-linguistic multimodal datasets, LAION-400M and LAION 2B-en, and models trained on them. We found evidence of hateful, aggressive, and targeted content in the alt text audit and evidence of racist stereotyping and dehumanizing classification in the models, particularly towards Black men, all of which exacerbates %
with dataset size. We cannot stress the importance of open-source in audit endeavors such as ours, since any kind of quantitative and qualitative dataset exploration hinges upon access to the artifacts themselves. We are saddened to see an increasing number of ML organizations fail to provide access to their datasets and models, since we believe that this is an essential element to scientific advancement and a healthy, equitable, and innovative research community.

Today’s state-of-the-art visio-linguistic multimodal models are trained with massive carbon footprints, massive data-infrastructure and massive funding. These models %
are currently being deployed in real-world including in recommendation systems, information-retrieval systems, semantic search systems and image captioning systems, although as we have illustrated in this paper, they can fail at %
associating photographs of human faces with the description: \textit{“A photo of a human being”}. Given that such failures can result in dire consequences on real people, often those at the margins of society, we implore the research community as well as those developing and deploying these systems to carry out due diligence with rigorous audits of these models and training datasets and take necessary actions, including refraining from use in high-stake scenarios.

\section*{Acknowledgements}
We would like to thank André Brock, Ellen Rushe, Gary Marcus, Sasha Luccioni, and Thomas Laurent for the invaluable comments on an earlier version of the paper.

\bibliographystyle{acm}
\bibliography{scale}

\clearpage 
\appendix
\section{OpenCLIP checkpoint naming conventions}
\label{app:openclip_naming}
In Table~\ref{tab:openclip_variants}, \texttt{openai} refers to OpenAI's closed WIT-400 million samples dataset, \texttt{laion400m\_e31(32)} refers to the checkpoint of the model trained on LAION-400M check-pointed after 31(32) training epochs, \texttt{s32b(s34b)} refers to checkpoints where the training was stopped after the model had \textit{seen} 32 (34) billion samples, \texttt{b79k(82k)(88k)} refers to training batch-sizes used (that is \texttt{batch\_size}=79000, 82000 and 88000 respectively). This information was not found in the documentation and was gleaned via a github issue raised. The conversation can be found here: \url{https://github.com/mlfoundations/open_clip/issues/454\#issuecomment-1451321921}. We also note that in the absence of a standardized model-naming template, it is hard to decipher details such as the batch size used for training certain models (especially those named \texttt{laion400m\_e31/e32}), which could potentially be another confounding parameter influencing the results obtained in this paper.
\section{The origins of the dataset scaling laws: A cartoon sketch emerges}
\label{app:dataset_scaling_laws}
While attempting to unearth what this specific dataset scaling law was that the practitioners were so inspired by, we repeatedly encountered a certain cartoon sketch 'power-law' plot referred to in both personal exchanges as well as in surveys such as~\cite{epoch2023scalinglawsliteraturereview}. As it turns out, this cartoon sketch power-plot first appeared as \texttt{Figure 6} in \textit{"Deep learning scaling is predictable, empirically"}~\cite{hestness2017deep}, a work that emerged out of Baidu research in 2017. The authors that first presented this plot posit that the generalization error associated with a ML  model exhibits a three phase behavior with regards to its training dataset size. The first phase, they state maps to  the 'small data region', where \textit{"models will struggle to learn from a small number of training samples"} resulting in high generalization errors. The second phase (or the middle portion of learning curves), they claim is the 'power-law region', where the generalization error monotonically decreases with training dataset size (linear with application-specific slopes when plotted on a log-log scale). This phase stretches till we hit of point of the 'glass-ceiling' or 'unbreachable error-floor' on account of factors such as model mismatch and mislabeled data (constitfuting the third phase). This, of course, has been further supplanted by the likes of the Chinchilla scaling laws (20 tokens per model parameter)~\cite{hoffmann2022training_chinchilla} in the specialized context of LLMs.

\section{Blackbox non-reproducible empirical results}
\label{app:dataset_scaling_results}
As for the blackbox non-reproducible empirical results that validated the dataset-scaling mandate and championed the \textit{scale-beats-noise} narrative, we refer to the ALIGN paper~\cite{jia2021scaling_align} that emerged in 2021. In the abstract section of this paper, we first encounter the following claim: \textit{“We show that the scale of our corpus can make up for its noise and leads to state-of-the-art representations even with such a simple learning scheme“}. The demonstration of this claim appears later in \texttt{"Section 6.2. Pre-training Datasets"} where the authors state that \textit{“To understand better how data size scaling wins over the increased noise, we further randomly sample 3M, 6M, and 12M ALIGN training data and compare them with the cleaned CC-3M data on B7+BERT-base model. Table 10 (sic) shows that while the ALIGN data performs much worse than CC data with the same size (3M), the model quality trained on 6M and 12M ALIGN data rapidly catches up. Despite being noisy, ALIGN data outperforms Conceptual Captions \textbf{with only 4x size}.”}
We note that these experiments (or similar ones) have not been replicated elsewhere to check if these scaling-ratios presented \textit{ipse dixit} in these contexts indeed hold true at all.

\section{The tactical template: Fuzzy main section meets non-existent appendices}
\label{app:tactical_template}
\begin{figure}
    \centering
    \includegraphics[width=\textwidth]{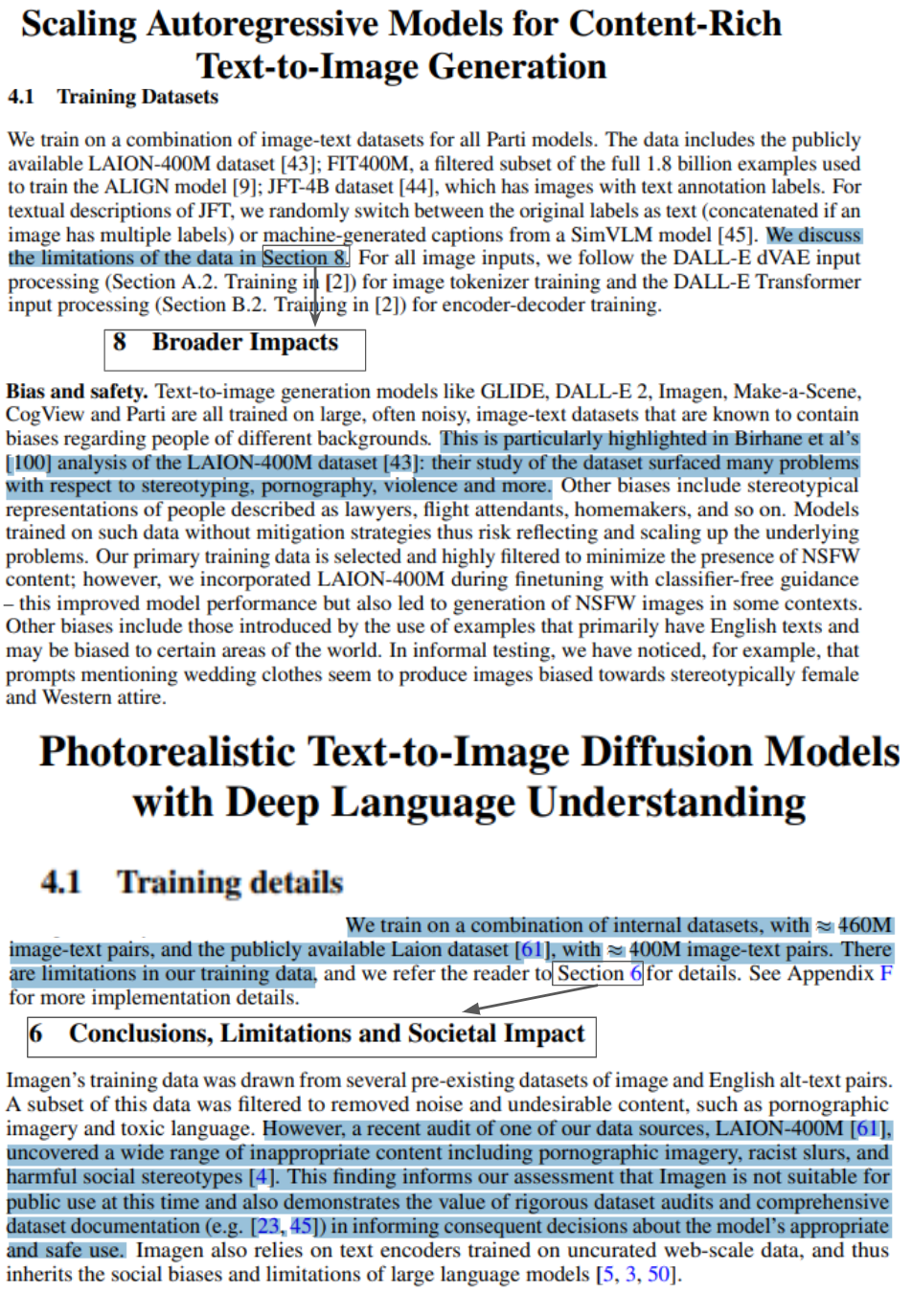}
    \caption{The Google template used to (non)declare the training dataset information along with paper screenshots}
\label{fig:parti_imagen}
\end{figure}
What unites the marquee projects of Dall-E, Parti and Imagen is the near-same tactical template deployed when it comes to (non)declaring the training dataset information. The template runs something like this:
\\ \textbf{Step-1:} Allocate a small nondescript subsection of the main section of the paper covering only the bare minimum details about the number of samples in the training dataset with cross-references to other similar blackbox datasets such as JFT. This coincidentally happens to be Section 4.1 in both Parti and Imagen papers (See Figure~\ref{fig:parti_imagen}).
\\ \textbf{Step-2: }Declare that somewhere in the succeeding sections titled on the lines of broader impacts or societal impacts, are details about the 'potentially problematic' aspects of the dataset and the downstream risks while patronisingly citing previously published audit papers (such as \cite{birhane2021multimodal} that have actually done the grunt work of exposing the gory details of such datasets. This happens to be Section 8 - Broader impacts in Parti and Section 6 for the Imagen model.
\\ \textbf{Step-3:} Setting the reader up for a non-existent Appendix section that is not part of the main-paper and not containing any details about how the dataset is actually constructed and where the data is sourced from  while noting the fact that its not mandatory for the reviewers to even glance at the Appendix section in peer-reviewed avenues of publishing.

It is in this backdrop we worryingly observe that the authors of the BASIC model paper have not even addressed model safety and dataset auditing issues in spite of having trained their model on the largest image-text dataset ever assembled and presented a full length 47 page paper with 3 revisions on ArXiv (See \url{https://arxiv.org/abs/2111.10050}).

\section{ The Vit-L/14 OpenCLIP Network Architecture}
\label{app:Vit-l-14}
The ViT-L/14 version of OpenCLIP has 428M parameters and 97M activations. Both the image and text branches output 768-dimensional embeddings. The image branch takes in images of size $224 \times 224$, and has a depth of 24 layers and a hidden dimension of size 1024. On the other hand, the text branch has a depth of 12 layers and a hidden dimension of size 768.

The ViT-L/14 version of OpenCLIP was trained using 400M samples from the LAION-400M dataset with a batch size of 96 per GPU over 400 GPUs for a total batch size of 38400 for 32 epochs. The learning rate was set to $6\times 10^{-4}$ with 5000 warm-up iterations. The total training time for the model was 88 hours.

\section{On AllLookSameism, negative stereotypes and racial misclassification}
\label{app:rg_exp}
\begin{figure*}[!ht]
    \centering
    \includegraphics[width=\textwidth]{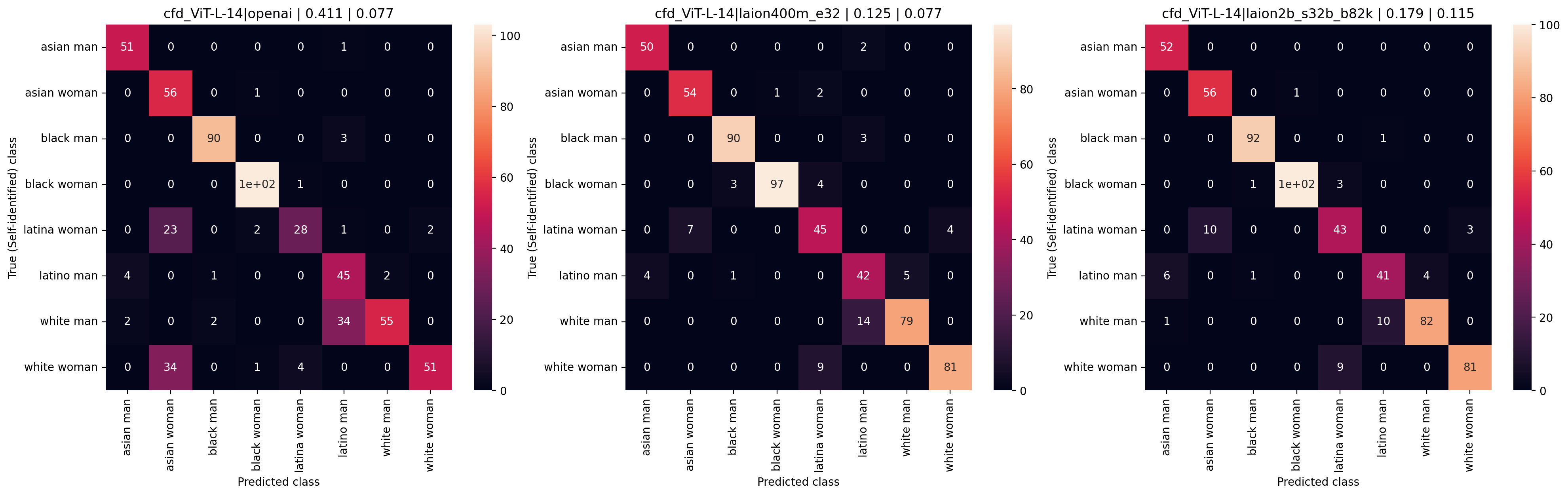}
    \caption{Heatmap of the confusion matrix of the race-gender classification experiment showing misclassification Latino/Latina individuals as `Asian' class. This misclassification got worse with dataset scaling.}
    \label{fig:crosstab_rg}
\end{figure*}
The goal here was to understand how stereotypes about facial appearances are cross-related with racial identities. When we looked that the results (Figure~\ref{fig:crosstab_rg}) we saw an interesting theme emerge: That the self-identified Latino/Latina individuals were misclassified with high confidence as one of the `Asian' classes on account of the presence of \textit{epicanthic folds} and this tendency to stereotype got worse with dataset scaling.
The title of these subplots here are formatted as strings with 4 fields separated by the `$|$' character:
$<\text{cfd\_Vit-L-14}>|<\text{training-dataset}>|<P_{lf \rightarrow af}>|<P_{lm \rightarrow am}>$. Here, $P_{lf \rightarrow af}$ is the probability that an image belongs to the \texttt{Latina-Female} category was misclassified as \texttt{Asian-Female} ( and $P_{lm \rightarrow am}$ is the probability that an image belong to the \texttt{Latina-Male} category was misclassified as \texttt{Asian-Male}). As seen in the first of the 3 subplots (from left) that maps to the OpenAI-WIT dataset 23 of the 56 latina women were misclassified as asian women leading to a $P_{lf \rightarrow af}=23/56=0.411$. This misclassification rate was better for the LAION-400M model ($0.125$) and worsened to $0.179$ for the LAION-2B-En model, thereby yielding yet another example of worsening of the bias-related metrics upon scaling the dataset from 400M to 2B samples. The same trend also showed up for Latino men with the misclassification rate increasingly nearly $50\%$ from $0.077$ to $0.115$.
\\
Correspondingly, there exists a substantial body of scientific literature (See \cite{fry2017latinx,gross2009own_latinx, pacheco2008rhetoric_latinx,nakamura2005alllooksame_latinx}) on not just the oft-ignored high levels of prevalence of the epicanthic folds in Hispanic/LatinX populations\footnote{``In Latinos, the inner canthal distance and lateral canthal angle of inclination were similar to Asians, while the lid crease spanned the range from Asians to Caucasians. Half of the Latinos had epicanthal folds"~\cite{fry2017latinx}} but also on the sociological ramifications of this \textit{alllooksame-ism}~\cite{nakamura2005alllooksame_latinx} that permeates aspects of the mainstream culture.

\section{ `Gorilla' in the alt-text descriptions}
\label{app:gorilla}

In this appendix, we provide details pertaining to the six images constituting the collage in Figure~\ref{fig:gorill}. The URLs and the Alt-text descriptions have not been edited or sanitized and are presented verbatim as we found them in the dataset.

\begin{enumerate}
    \item Sub-figure (a) was sourced from \url{http://pigroll.com/img/youre_gonna_get_raped.jpg} with the alt-text description: \texttt{"Smiling albino gorilla wants to rape you"}
    \item Sub-figure (b) was sourced from \url{https://content.wafflegirl.com/galleries/content/2/247/2247763_ec5249d.jpg} with the alt-text description: \texttt{"men fuck gorilla girl"}.
    \item Sub-figure (c) was sourced from \url{https://media.istockphoto.com/photos/female-gorilla-with-baby-picture-id578309582} with the alt-text description: \texttt{"Woman stripper with gorilla partner, girl fuck with eel"}
    \item Sub-figure (d) was sourced from \url{https://new4.fjcdn.com/pictures/Gorilla+warfare_d41bd9_3543352.jpg} with the alt-text description: \texttt{"Gorilla warfare. . Top Comments This has been Bagged as more hide r Not Spam What the fuck did you just say about me, you little bitch? I' ll have you know I gr Gorilla warfare Top Comments This has been Bagged as more hide r Not Spam What the fuck did you just say about me little bitch? I' ll have know I gr"}
    \item Sub-figure (e) was sourced from \url{https://farm1.static.flickr.com/552/31262527324_caa242e8d8_m.jpg}  with the alt-text description: \texttt{"Mama Ayesha's Restaurant and Donald Trump (The Devils in the Details) Tags: donaldtrump mamaayeshas wallofpresidents hanksywashingtondc cia gop isis vladimirputin russia sexdrugsandrockandroll hillaryclinton plannedparenthood bigot dumptrump thewalkingdead republican pedophile mikepence nastywoman badhombre conservative rape riencepriebus donaldmcgahn stevenbannon frankgaffney jeffsessions generaljamesmattis generaljohnkelly stevenmnuchin andypuzder wilburross cathymcmorrisrodgers trumpforpresidentbobblehead poopydiaper ktmcfarland mikepompeo nikkihaley betsydevos tomprice scottpruitt seemaverma gorilla marriageequality kukluxklan daryldixon newyorkcity melaniatrump riggedelection jihad terrorist taliban mexicanwall racism confederateflag nazi islam freedom berniesanders americannaziparty thebeatles therollingstones democrat civilrights tednugent tempertantrum contraception abortion tinfoilhatsociety michelleobama she'sacunt foxnews liberal"}
    \item Sub-figure (f) was sourced from \url{https://see.xxx/mt/sL/1994633.jpg} with the alt-text description: \texttt{"Hirsute wet crack of this gorilla lady is so nasty that dont crave to fuck that"}. 
\end{enumerate}

\end{document}